\documentclass[%
 aip,
 amsmath,amssymb,
 reprint,%
]{revtex4-1}

\usepackage{graphicx}
\usepackage{dcolumn}
\usepackage{bm}

\usepackage[utf8]{inputenc}
\usepackage[T1]{fontenc}
\usepackage{mathptmx}
\usepackage{etoolbox}
\usepackage{sidecap}
\usepackage{booktabs}
\usepackage{array}
\usepackage{soulutf8}
\usepackage{colortbl}
\usepackage{xcolor}

\sethlcolor{yellow}

\makeatletter
\def\@email#1#2{%
 \endgroup
 \patchcmd{\titleblock@produce}
  {\frontmatter@RRAPformat}
  {\frontmatter@RRAPformat{\produce@RRAP{*#1\href{mailto:#2}{#2}}}\frontmatter@RRAPformat}
  {}{}
}%
\makeatother
\begin{document}
\setlength{\parskip}{5pt}
\preprint{AIP/123-QED}

\title{Development of Anisotropic Magnetized Viscosity for Magnetized Liner Inertial Fusion Simulations in FLASH}
\author{A. Sam}
\altaffiliation{ashwyn.sam@pacificfusion.com}

\affiliation{ 
Department of Aeronautics and Astronautics, Stanford University, 496 Lomita Mall Lane, Stanford, CA 94305, USA
}%
\affiliation{ 
Pacific Fusion Corporation, 6082 Stewart Ave., Fremont, CA 94538, USA
}%

\author{F. Garcia-Rubio}%
\author{S. Davidson}%
\author{C.~L. Ellison}%
\author{J. Hamilton}%
\affiliation{ 
Pacific Fusion Corporation, 6082 Stewart Ave., Fremont, CA 94538, USA
}%
\author{R. Lau}%
\affiliation{ 
Department of Aeronautics and Astronautics, Stanford University, 496 Lomita Mall Lane, Stanford, CA 94305, USA
}%
\affiliation{ 
Pacific Fusion Corporation, 6082 Stewart Ave., Fremont, CA 94538, USA
}%

\author{N. Meezan}%
\author{A. Reyes}%
\author{P. Schmit}%
\author{A. L. Velikovich}%
\affiliation{ 
Pacific Fusion Corporation, 6082 Stewart Ave., Fremont, CA 94538, USA
}%

\begin{center}
\begin{minipage}{0.8\textwidth}
\begin{abstract}
Magnetized liner inertial fusion (MagLIF) operates in a regime where anisotropic transport phenomena fundamentally influence implosion dynamics. In strongly magnetized plasmas, the viscous stress tensor becomes highly anisotropic, yet no prior work has incorporated or examined magnetized viscosity effects in MagLIF configurations. We present the first implementation of the full Braginskii magnetized viscosity tensor for arbitrary magnetic field orientations in the Pacific Fusion branch of FLASH. The implementation is verified through analytical comparisons, direct verification against Braginskii's original formulation, Method of Manufactured Solutions, and against analytical shock solutions. Application to MagLIF-relevant configurations reveals that magnetized viscosity damps vortical structures, converts kinetic energy in those vortical structures into thermal energy, and mitigates the Rayleigh-Taylor instabilities. Simulations with seeded perturbations demonstrate yield preservation when magnetized viscosity is included. These results establish magnetized viscosity as a non-negligible physical mechanism in MagLIF plasmas and provide a validated capability for predictive modeling of magnetized high-energy-density plasmas.
\end{abstract}
\end{minipage}
\end{center}

\maketitle

\section{\label{sec:level1}Introduction}

Pulser ICF concepts, such as magnetized liner inertial fusion (MagLIF), represent a promising pathway toward controlled thermonuclear fusion.\cite{gomez_magnetized_2025} In MagLIF, a cylindrical liner implodes onto a magnetized, preheated fuel column, compressing both plasma and magnetic field to fusion-relevant conditions. This approach combines three key elements: preheating of the fuel, axial magnetic field compression, and pulsed-power driven implosion to achieve the conditions necessary for fusion ignition. The MagLIF concept operates in an intermediate regime between magnetic confinement fusion and traditional inertial confinement fusion, with typical parameters including densities of $\sim$10$^{23}$ cm$^{-3}$, and confinement times of $\sim$10$^{-9}$ s. Recent experiments on the Z facility have demonstrated fusion-relevant ion temperatures up to 3.1 keV and thermonuclear production of up to 1.1 $\times$ 10$^{13}$ deuterium-deuterium neutrons. \cite{gomez_performance_2020} The platform development has enabled increases in applied magnetic field from 10 to 20 T through internally reinforced coil designs, coupled preheat energy from less than 1 to 2.3 kJ through improved laser pulse shaping and beam smoothing, and peak load current from 16 to 20 MA through redesign of the final transmission line.\cite{gomez_magnetized_2025} Pulser ICF systems provide several advantages over laser-driven ICF, including lower costs, better economies of scale, improved efficiency, and longer operational lifetimes. \cite{ellison_opportunities_2025} These characteristics have motivated recent investigations into MagLIF as a potential source for fusion power plant operations.

While conduction losses are important in both laser-driven and magnetically driven ICF, the relative importance of thermal conduction versus radiative losses differs significantly between the two approaches. Laser-ICF schemes typically involve higher densities and areal densities ($\rho R$) in the hot spot, which shifts the energy loss balance toward radiation. In MagLIF, with characteristic stagnation temperatures of $\sim$4 keV and $\rho R \sim 0.01$ g/cm$^2$, the ratio of electron thermal conduction losses to bremsstrahlung radiation losses is approximately 64, making conduction the dominant loss mechanism.\cite{schmit_conservative_2020} Without magnetic confinement to suppress perpendicular heat conduction, thermal energy would be rapidly lost from the fuel to the liner, presenting a fundamental challenge that magnetization directly addresses.

Viscosity, often disregarded as a transport term, may play an important role in MagLIF plasmas by damping hydrodynamic instabilities, shearing axial flows resulting from non-homogeneous preheat, and dissipating fuel kinetic energy in vortical structures into thermal energy.\cite{davidovits_viscous_2019} Near stagnation, strong velocity gradients could develop where viscous heating may contribute to the energy budget. In the strongly magnetized regime characteristic of MagLIF plasmas, strong magnetic fields fundamentally alter the viscous stress tensor through ion particle gyromotion, reduces perpendicular viscosity while leaving parallel viscosity unaffected, thereby creating preferential momentum transport along field lines. Recent 3D magnetohydrodynamics (MHD) simulations of MagLIF implosions have identified several areas where the absence of magnetized viscosity limits predictive capability.\cite{weis_assessing_2025} Discrepancies between simulated and experimentally inferred stagnation temperatures of $\sim$10--15\% could potentially be explained by viscous dissipation of azimuthal kinetic energy in the fuel. Additionally, simulations may overestimate the effects of laser-driven azimuthal flows near the fuel-liner interface due to the lack of viscosity that could dissipate this kinetic energy. Previous high-resolution simulations of National Ignition Facility (NIF) capsules have shown that including physical viscosity reduces hot-spot Reynolds numbers to $\mathrm{Re} \sim 10\text{--}100$ and strongly damps small-scale turbulent motions~\cite{Weber2014HotSpotViscosity}. These observations motivate the present work: a complete implementation of the Braginskii magnetized viscosity tensor that enables direct assessment of viscous effects on MagLIF performance.

Despite the potential importance of anisotropic viscosity in magnetized high-energy-density plasmas, there has been no prior work examining Braginskii viscosity effects in MagLIF configurations. \cite{weis_assessing_2025} This work presents the first implementation of Braginskii's magnetized viscosity formulation for an arbitrary magnetic field into the Pacific Fusion branch of FLASH \cite{fryxell_flash_2000, ellison_validation_2025} to enable high-fidelity simulations of MagLIF-like plasmas with accurate treatment of anisotropic transport phenomena.\cite{braginskii_transport_1965} 

Given the absence of MagLIF-specific studies, we draw from intracluster medium (ICM) research where Braginskii viscosity profoundly affects plasma dynamics. The weakly collisional ICM ($T \sim 1$--$10$ keV, $n \sim 10^{-2}$ cm$^{-3}$, $L \gg \lambda_{\text{mfp}} \gg r_i \gg r_e$, where $L$ is the system scale length, $\lambda_{\text{mfp}}$ is the mean free path, and $r_i$ and $r_e$ are the ion and electron gyroradii, respectively)\cite{vikhlinin_chandra_2006,schekochihin_turbulence_2006} exhibits anisotropic transport that invalidates classical stability criteria,\cite{schwarzschild_structure_1959,balbus_backward_2000,balbus_convective_2001,quataert_buoyancy_2008} driving the magnetothermal instability (MTI) and heat-flux-driven buoyancy instability (HBI).\cite{balbus_backward_2000,balbus_convective_2001,quataert_buoyancy_2008} Braginskii viscosity critically modifies these instabilities, causing HBI wavelengths to exceed atmospheric scale heights,\cite{kunz_dynamical_2011,latter_hbi_2012} preventing complete field reorientation, and providing viscous heating.\cite{kunz_buoyancy_2012} It suppresses Kelvin-Helmholtz instability,\cite{zuhone_effect_2014,suzuki_magnetohydrodynamic_2013} stabilizes active galactic nuclei (AGN) bubbles,\cite{dong_buoyant_2009} reduces mixing,\cite{parrish_effects_2012,berlok_local_2016,berlok_helium_2016} and damps magnetosonic waves while preserving Alfv\'en waves.\cite{parrish_effects_2012,squire_stringent_2016,squire_amplitude_2017} In MagLIF, perpendicular viscosity could similarly suppress velocity shear along field lines, potentially stabilizing magneto-Rayleigh-Taylor modes and improving flux conservation, while viscous heating from converging flow provides additional thermal energy at stagnation.

The influence of magnetized viscosity on shock structures provides another important context for understanding Braginskii transport. Viscous dissipation converts a substantial fraction of the incoming kinetic energy into internal energy, with the Reynolds number across the shock front remaining of order unity; this fundamental requirement sets the characteristic shock thickness.\cite{LandauLifshitz_FM,ZeldovichRaizer_Shock} For propagation perpendicular to $\boldsymbol{B}$, the viscous-stress components associated with shear are modified by the ion Hall parameter $\omega_i \tau_i$, where $\omega_i$ is the ion cyclotron frequency and $\tau_i$ is the ion-ion collision time: the perpendicular (cross-field) shear viscosity decreases roughly as $\eta_\perp \sim \eta_0\,[1+(\omega_i \tau_i)^2]^{-1}$, while the gyroviscous terms decrease as $\eta_{\wedge} \sim \eta_0\,\omega_i\tau_i\,[1+(\omega_i \tau_i)^2]^{-1}$. Thus, in the strongly magnetized limit ($\omega_i \tau_i \gg 1$), cross-field momentum transport is suppressed.\cite{braginskii_transport_1965, hollweg_viscosity_1985, kaufman_plasma_1960} This anisotropy modifies the internal shock structure: because the ion viscosity coefficient scales as $\eta_0 \propto T_i^{5/2}$, heating within the front tends to broaden a viscous shock, whereas magnetization reduces the effective cross-field viscosity and therefore tends to narrow the transition. The resulting shock thickness reflects the competition between these two effects.\cite{LandauLifshitz_FM,ZeldovichRaizer_Shock} These competing mechanisms alter the dissipation profile and the spatial structure of the shock in both astrophysical and laboratory plasmas; numerically capturing them requires an anisotropic (Braginskii) viscosity model rather than an isotropic Navier--Stokes closure.\cite{berlok_braginskii_2020} In MagLIF-relevant conditions, where shocks may arise during fuel compression and transport is strongly shaped by magnetization, these effects could influence the detailed structure of shock-mediated dissipation and are therefore relevant to high-fidelity modeling.\cite{slutz_pulsed-power-driven_2010, gomez_experimental_2014, slutz_scaling_2016}

Viscosity has long been recognized as a stabilization mechanism for hydrodynamic instabilities, yet its effectiveness varies dramatically across different plasma regimes. In classical Rayleigh-Taylor theory, viscous forces can substantially reduce instability growth rates, with the damping effect characterized by the competition between buoyancy and viscous dissipation.\cite{chandrasekhar_hydrodynamic_2013} The magneto-Rayleigh-Taylor instability (MRTI) poses a particular challenge for MagLIF implosions: during compression, magnetic pressure drives the dense liner inward with extreme accelerations, creating conditions analogous to a light fluid accelerating into a heavy fluid.\cite{mcbride_beryllium_2013, slutz_pulsed-power-driven_2010, sinars_measurements_2010, sinars_measurements_2011} Under these conditions, viscous stabilization becomes exceptionally difficult to achieve. The situation reverses during the deceleration phase, where the inner liner and dense fuel layer accelerate into the lower-density hot spot, making this interface unstable. Recent theoretical work has shown that viscous MRTI growth depends on a single dimensionless parameter, the Galilei number Ga, which measures the ratio of gravitational to viscous forces.\cite{dai_linear_2023, lau_impact_2026} For acceleration-phase MRTI in the solid liner, achieving meaningful viscous stabilization requires viscosities exceeding $10^2$ g/cm$\cdot$s even at modest drive currents,\cite{lau_impact_2026} making viscous damping of these modes practically unattainable. However, the situation may be more favorable for deceleration-phase MRTI at the fuel-liner interface, where the plasma is much hotter and less dense compared to the liner material. In this regime, viscosity could potentially reduce the deleterious effects of deceleration-phase instabilities on hot spot integrity. This possibility underscores why accurate modeling of magnetized transport, including both temperature-dependent viscosity enhancement and magnetic field-induced anisotropy, remains essential for understanding instability evolution in MagLIF implosions.

Braginskii viscosity contributes directly to the internal--energy balance via the work term
\(Q_{\mathrm{visc}}=-\boldsymbol{\Pi}:\nabla\mathbf{v}\), providing an anisotropic pathway that preferentially converts field-aligned shear into thermal energy. In MagLIF-like compressions, nonradial flow represents ``wasted'' compression energy unless it is viscously dissipated. \citet{davidovits_viscous_2019} showed for 2-D compressions that, with adiabatic heating and the strong Braginskii scaling \(\mu\propto T^{5/2}\), shear and turbulent motions can be efficiently converted into heat during compression, although complete viscous dissipation generally requires sufficiently large compression and depends on the mode structure and boundary conditions. This motivates including magnetized anisotropic viscosity in simulations to quantify the extent to which viscous heating near stagnation modifies the fuel energy budget alongside thermal conduction and radiation in quasi-isobaric MagLIF implosions.\cite{davidovits_viscous_2019}

This work focuses on ion viscosity within the Braginskii transport closure. In the MHD framework considered here, bulk electron motion is not resolved and electron inertia is neglected; consequently, the electron stress tensor enters only as a higher-order correction to the momentum and energy equations, smaller than the ion contribution by a factor of $m_e/m_i \approx 1/1836$ (for $Z = 1$), and is therefore neglected. Although substantial effort has been devoted to refining Braginskii’s original electron transport coefficients \cite{epperlein_plasma_1986, davies_transport_2021, simakov_electron_2022}, including extensions to higher-$Z$ plasmas where electron viscosity can become significant \cite{simakov_electron_2022, whitney_momentum_1999, velikovich_role_2001, miller_splitting_2020, zhang_influence_2024}, these developments pertain to the electron species and are outside the scope of the present study. For ions, Braginskii’s original formulation remains the standard description of collisional, magnetized plasma viscosity. Comparison against this formulation is essential because a large body of existing literature and simulation capability is based on Braginskii’s viscosity formulas.


This work presents the first implementation of the full Braginskii viscosity tensor in an implicit solver within a multiphysics radiation-hydrodynamics framework. The viscosity tensor affects both momentum transport through the divergence of the anisotropic stress and energy balance through viscous heating. Our implicit treatment of velocity diffusion offers two critical advantages: it eliminates restrictive viscous CFL constraints that would otherwise dominate timesteps in high-viscosity regions, and it maintains numerical stability across the extreme viscosity variations—spanning many orders of magnitude—that occur between cold liner material and hot fusion fuel. This implementation represents a continuation of Pacific Fusion's comprehensive modeling efforts to develop high-fidelity simulation capabilities for MagLIF target design. \cite{ellison_opportunities_2025, ellison_validation_2025, alexander_affordable_2025, garcia-rubio_analysis_2025, farmer_numerical_2024}

The remainder of this paper is organized as follows. Section 2 presents the mathematical formulation of the Braginskii viscosity tensor and its incorporation into the MHD equations. Section 3 details the numerical implementation. Section 4 verifies the implementation through comparison with analytical solutions and benchmark problems. Section 5 demonstrates the impact of magnetized viscosity on MagLIF-relevant test problems. Finally, Section 6 discusses implications for MagLIF performance and future extensions.

This work represents an essential step toward predictive modeling of magnetized high-energy-density plasmas. By capturing the full anisotropic transport physics, these simulations can now assess how magnetized viscosity influences shock heating, instability growth, magnetic flux evolution, and energy confinement throughout the implosion, ultimately enabling more accurate predictions of fusion performance in MagLIF and related magnetized ICF concepts.

\section{Mathematical Formulation}
\label{sec:math}

The viscous contribution to the momentum equation enters through the stress divergence term,
\begin{equation}
  \rho\,\frac{\partial \mathbf{v}}{\partial t}\bigg|_{\text{visc}}
  = -\nabla\cdot\boldsymbol{\Pi},
  \label{eq:visc_vel_update}
\end{equation}
where $\rho$ is the density, $\mathbf{v}$ is the velocity vector, $t$ is time, and $\boldsymbol{\Pi}$ denotes the viscous stress tensor. The corresponding viscous heating contribution to the energy equation takes the form
\begin{equation}
  Q_{\mathrm{visc}} = -\boldsymbol{\Pi}:\nabla\mathbf{v}.
  \label{eq:Qvisc_vector}
\end{equation}
For standard hydrodynamics, the stress tensor is an isotropic quantity given by $\boldsymbol{\Pi} = -\eta_0 \mathbf{W}$, where $\eta_0$ is the viscosity coefficient and $\mathbf{W}$ is the symmetric, traceless rate-of-strain tensor defined as
\begin{equation}
  \mathbf{W} = \nabla\mathbf{v} + (\nabla\mathbf{v})^T - \frac{2}{3}(\nabla \cdot \mathbf{v})\mathbf{I},
  \label{eq:Wdef_vector}
\end{equation}
where $\mathbf{I}$ is the identity tensor.

In the presence of a magnetic field, momentum transport becomes fundamentally anisotropic, with distinct behavior parallel and perpendicular to the field lines. Furthermore, the direction of the transported momentum relative to the velocity gradient becomes physically significant. Consequently, the constitutive relation between the viscous stress tensor $\boldsymbol{\Pi}$ and the rate-of-strain tensor $\mathbf{W}$ becomes substantially more complex, requiring five independent viscosity coefficients. This number arises naturally from the fact that a symmetric, traceless tensor possesses exactly five independent components; therefore, the most general linear relationship between two such tensors requires five independent coefficients of proportionality.\cite{braginskii_transport_1965}

Following Braginskii's formulation (and adopting his Einstein index notation) \cite{braginskii_transport_1965}, the anisotropic viscous stress tensor in a magnetized plasma takes the form
\begin{equation}
  \Pi_{\alpha\beta}
  =
  -\eta_0 W_{0\,\alpha\beta}
  -\eta_1 W_{1\,\alpha\beta}
  -\eta_2 W_{2\,\alpha\beta}
  +\eta_3 W_{3\,\alpha\beta}
  +\eta_4 W_{4\,\alpha\beta},
  \label{eq:pi_decomp}
\end{equation}
where the five basis tensors $W_{k\,\alpha\beta}$ form a mutually orthogonal set constructed from the rate-of-strain tensor $W_{\mu\nu}$ and the magnetic field direction. Here the spatial indices $\alpha, \beta, \mu, \nu$ correspond to $(x,y,z)$ in Cartesian coordinates or $(r,\theta,z)$ in cylindrical coordinates. The implementation presented in this work adopts the cylindrical geometry framework. These basis tensors are expressed in terms of the Kronecker delta $\delta_{\alpha\beta}$, the Levi-Civita tensor $\varepsilon_{\alpha\beta\gamma}$, and the unit vector along the magnetic field $\mathbf{h} \equiv \mathbf{B}/B$ with components $h_\alpha$:
\begin{subequations}
\label{eq:Wk}
\begin{align}
  W_{0\,\alpha\beta} 
  &= \frac{3}{2}\left(h_\alpha h_\beta - \frac{1}{3}\delta_{\alpha\beta}\right)
     \left(h_\mu h_\nu - \frac{1}{3}\delta_{\mu\nu}\right) W_{\mu\nu},
  \\
  W_{1\,\alpha\beta} 
  &= \left(\delta^{\perp}_{\alpha\mu}\,\delta^{\perp}_{\beta\nu} + \frac{1}{2}\delta_{\alpha\beta}\,h_\mu h_\nu\right) W_{\mu\nu},
  \\
  W_{2\,\alpha\beta} 
  &= \left(\delta^{\perp}_{\alpha\mu} h_\beta h_\nu + \delta^{\perp}_{\beta\nu} h_\alpha h_\mu\right) W_{\mu\nu},
  \\
  W_{3\,\alpha\beta} 
  &= \frac{1}{2}\left(\delta^{\perp}_{\alpha\mu}\,\varepsilon_{\beta\nu\gamma}
     + \delta^{\perp}_{\beta\nu}\,\varepsilon_{\alpha\mu\gamma}\right) h_\gamma W_{\mu\nu},
  \\
  W_{4\,\alpha\beta} 
  &= \left(h_\alpha h_\mu \varepsilon_{\beta\nu\gamma} + h_\beta h_\nu \varepsilon_{\alpha\mu\gamma}\right) h_\gamma W_{\mu\nu},
\end{align}
\end{subequations}
where $\delta^{\perp}_{\alpha\beta} \equiv \delta_{\alpha\beta} - h_\alpha h_\beta$ is the perpendicular projection operator.

The five viscosity coefficients, $\eta_k = \eta_k(n_i, T_i, \omega_i \tau_i)$, depend on the local plasma state, specifically the ion number density $n_i$ and temperature $T_i$, as well as the ion Hall parameter $\omega_i \tau_i \equiv (ZeB/m_ic)\tau_i$, where $Z$ is the ion charge, $B$ is the magnetic-field strength, $m_i$ is the ion mass, and $\tau_i$ is the ion–ion collision time. Their explicit forms are given in Eq. (4.44) of Braginskii \cite{braginskii_transport_1965}. 

Each coefficient governs distinct momentum transport processes: $\eta_0$ controls field-parallel compression and stretching in tandem with the isotropic component of the compression perpendicular to the magnetic field; $\eta_1$ handles stresses formed entirely within the plane perpendicular to the magnetic field (e.g., for $\mathbf{B} \parallel \hat{z}$, this includes rate-of-strain components like $W_{xx} - W_{yy}$, and $W_{xy}$); $\eta_2$ governs shear stresses involving both parallel and perpendicular directions (e.g., $W_{xz}$ and $W_{yz}$), coupling field-aligned and cross-field motions; while $\eta_3$ and $\eta_4$ represent gyroviscous coefficients arising from finite ion Larmor radius effects. These gyroviscous terms contain cross-product structures that enable momentum transfer in directions perpendicular to both the applied strain and the magnetic field, analogous to Hall-type transport, redistributing momentum without dissipation.

In the high magnetization limit where $\omega_i\tau_i \gg 1$, shear stresses and the gyroviscous terms become severely restricted, scaling as  $(\omega_i\tau_i)^{-2}$ and $(\omega_i\tau_i)^{-1}$, respectively, whereas isotropic compression, governed by $\eta_0$, maintains the same magnitude as in the unmagnetized case.

The viscous heating contribution to the energy equation is shown in (\ref{eq:Qvisc_vector}) where notably the gyroviscous terms ($\eta_3, \eta_4$) are not included due to their non-dissipative nature. The heating rate remains finite and positive. In strongly magnetized plasmas, viscous heating predominantly arises from field-aligned motions, as the perpendicular dissipative coefficients become strongly suppressed.

\section{Numerical Implementation}
\label{sec:numerical}

The implementation of the Braginskii viscosity tensor in the Pacific Fusion branch of FLASH requires careful treatment of the anisotropic momentum diffusion arising from the magnetized transport coefficients. This section describes the numerical approach, focusing on the implicit treatment necessary to maintain stability across the wide range of viscosity values encountered in MagLIF simulations.

The viscous contribution to the momentum equation \eqref{eq:visc_vel_update} is discretized using an implicit scheme to avoid restrictive timestep constraints. The discrete momentum update takes the form
\begin{equation}
  \rho \frac{v_i^{n+1} - v_i^n}{\Delta t} = -\frac{1}{r}\frac{\partial}{\partial j}\left(r\Pi_{ij}^{n+1}\right),
  \label{eq:implicit_momentum}
\end{equation}
where $v_i^n$ denotes the velocity component at time level $n$, $\Delta t$ is the timestep, and the factor of $r$ accounts for cylindrical geometry. The superscript $n+1$ on the stress tensor indicates implicit evaluation at the new time level. In our 2D cylindrical simulations, we update velocities in the radial, azimuthal, and axial directions but maintain axisymmetry (i.e., $\partial v_i/\partial \theta = 0$).

\subsection{HYPRE Interface}

To leverage the efficient parallel solvers in the HYPRE library, we reformulate equation \eqref{eq:implicit_momentum} into the canonical form solved by HYPRE's structured grid interface \cite{FLASH_Users_Guide_4p8}:
\begin{align}
  \rho \frac{\partial v_i}{\partial t} = &\frac{1}{r}\frac{\partial}{\partial k}\left(r B_{ijkl} \frac{\partial v_j}{\partial l} + E_{ijk} v_j\right) + \frac{1}{r}F_{ijk}\frac{\partial v_j}{\partial k} \nonumber\\
  &+ \frac{1}{r^2}G_{ij}v_j + \frac{\partial}{\partial k}\left(W_{ijkl}\frac{\partial v_j}{\partial l}\right),
  \label{eq:hypre_form}
\end{align}
where the coefficient tensors $B_{ijkl}$, $E_{ijk}$, $F_{ijk}$, $G_{ij}$, and $W_{ijkl}$ encode the anisotropic viscosity physics. We note that the rank-4 tensor $W_{ijkl}$ corresponds to the HYPRE formulation and should not be confused with the rank-3 basis tensor $W_{k\alpha\beta}$ defined earlier; the same symbol is retained here to remain consistent with both HYPRE and Braginskii notation.

The key challenge is to map from the Braginskii formulation to these HYPRE coefficients. This requires expressing the stress tensor divergence in a form where we can systematically identify terms according to their differential structure. The remainder of this section describes the tensor manipulations necessary to achieve this mapping.

\subsection{Stress Tensor Reformulation}

Substituting the Braginskii stress decomposition \eqref{eq:pi_decomp} and expressing each basis tensor $W_{k\,ij}$ in terms of velocity gradients yields a linear system for the updated velocity components. The key insight is that the stress tensor $\Pi_{ij}$ can be written as a linear operator acting on a generalized velocity gradient vector that includes both spatial derivatives and geometric terms arising from the cylindrical coordinate system. Specifically, we can express:
\begin{equation}
  \Pi_{\alpha\beta} = -\mathcal{L}_{\alpha\beta j} q_j,
  \label{eq:stress_linear_operator}
\end{equation}
where $q_j$ is a vector (shown in (\ref{eq:gradient_vector})) containing all velocity derivatives (e.g., $\partial_r v_r$, $\partial_z v_\theta$) as well as geometric terms (e.g., $v_r/r$, $v_\theta/r$), and $\mathcal{L}_{\alpha\beta j}$ are coefficient tensors that depend on the local magnetic field direction $\mathbf{h}$ and the five Braginskii viscosity coefficients. We use $j$ to represent the index that goes from 1 to 8. In the following paragraphs, we will use $\alpha, \beta, \mu, \nu$ to represent the spatial indices and use $i,j,k$ indices to represent any additional integer indices in the tensor.

To facilitate implementation in FLASH's existing framework and enable the mapping to HYPRE coefficients, we need to find $\mathcal{L}_{\alpha\beta j}$. This requires expressing the basis tensor $W_{\alpha\beta}$ in terms of a transformation matrix $T_{\mu\nu j}$ and a gradient vector $q_j$.

The velocity gradient tensor $W_{\alpha\beta}$ in cylindrical coordinates $(r, \theta, z)$ without assuming $v_\theta = 0$ is given by:
\begin{equation}
  W_{\alpha\beta} = 
  \begin{bmatrix}
    2\partial_r v_r - \frac{2}{3}\nabla \cdot \mathbf{v} & 
    -\frac{v_\theta}{r} + \partial_r v_\theta & 
    \partial_z v_r + \partial_r v_z \\[0.3em]
    -\frac{v_\theta}{r} + \partial_r v_\theta & 
    \frac{2v_r}{r} - \frac{2}{3}\nabla \cdot \mathbf{v} & 
    \partial_z v_\theta \\[0.3em]
    \partial_z v_r + \partial_r v_z & 
    \partial_z v_\theta & 
    2\partial_z v_z - \frac{2}{3}\nabla \cdot \mathbf{v}
  \end{bmatrix},
  \label{eq:W_full}
\end{equation}
where $\nabla \cdot \mathbf{v} = \partial_r v_r + v_r/r + \partial_z v_z$ denotes the velocity divergence in cylindrical coordinates.

We introduce a gradient vector containing all relevant velocity derivatives and geometric terms:
\begin{equation}
  \begin{bmatrix}
    q_1 \\ q_2 \\ q_3 \\ q_4 \\ q_5 \\ q_6 \\ q_7 \\ q_8
  \end{bmatrix}
  = 
  \begin{bmatrix}
    \partial_r v_r \\
    \partial_r v_\theta \\
    \partial_r v_z \\
    \partial_z v_r \\
    \partial_z v_\theta \\
    \partial_z v_z \\
    v_r/r \\
    v_\theta/r
  \end{bmatrix},
  \label{eq:gradient_vector}
\end{equation}

The transformation matrix $T_{\mu\nu j}$ relates the velocity gradients to the tensor components $W_{\alpha\beta}$ through
\begin{equation}
  W_{\mu\nu} = T_{\mu\nu j} q_j.
  \label{eq:W_transform}
\end{equation}
The nonzero elements of this transformation matrix are:
\begin{align}
  T_{111} &= \frac{4}{3}, \quad T_{221} = T_{331} = -\frac{2}{3}, \nonumber\\
  T_{122} &= T_{212} = 1, \nonumber\\
  T_{133} &= T_{313} = 1, \nonumber\\
  T_{134} &= T_{314} = 1, \nonumber\\
  T_{235} &= T_{325} = 1, \nonumber\\
  T_{116} &= T_{226} = -\frac{2}{3}, \quad T_{336} = \frac{4}{3}, \nonumber\\
  T_{117} &= T_{337} = -\frac{2}{3}, \quad T_{227} = \frac{4}{3}, \nonumber\\
  T_{128} &= T_{218} = -1,
  \label{eq:T_elements}
\end{align}
with all other $T_{\alpha\beta j} = 0$.

Following the Braginskii formulation, we introduce the rank-4 tensors $Q_{k\alpha\beta\mu\nu}$ that encode the geometric structure of each viscosity mode. These tensors are defined as:
\begin{align}
  Q_{0\alpha\beta\mu\nu} &= \frac{3}{2}\left(h_\alpha h_\beta - \frac{1}{3}\delta_{\alpha\beta}\right)\left(h_\mu h_\nu - \frac{1}{3}\delta_{\mu\nu}\right), \nonumber\\
  Q_{1\alpha\beta\mu\nu} &= \delta_{\alpha\mu}^\perp \delta_{\beta\nu}^\perp + \frac{1}{2}\delta_{\alpha\beta}^\perp h_\mu h_\nu, \nonumber\\
  Q_{2\alpha\beta\mu\nu} &= \delta_{\alpha\mu}^\perp h_\beta h_\nu + \delta_{\beta\nu}^\perp h_\alpha h_\mu, \nonumber\\
  Q_{3\alpha\beta\mu\nu} &= \frac{1}{2}\left(\delta_{\alpha\mu}^\perp \epsilon_{\beta\gamma\nu} + \delta_{\beta\nu}^\perp \epsilon_{\alpha\gamma\mu}\right) h_\gamma, \nonumber\\
  Q_{4\alpha\beta\mu\nu} &= \left(h_\alpha h_\mu \epsilon_{\beta\gamma\nu} + h_\beta h_\nu \epsilon_{\alpha\gamma\mu}\right) h_\gamma,
  \label{eq:Q_tensors}
\end{align}

The linear coefficient tensor $\mathcal{L}_{\alpha\beta j}$ is then constructed as:
\begin{equation}
  \mathcal{L}_{\alpha\beta j} = \eta_k Q_{k\alpha\beta\mu\nu} T_{\mu\nu j},
  \label{eq:P_construction}
\end{equation}
where $\eta_k = [\eta_0, \eta_1, \eta_2, -\eta_3, -\eta_4]$ are the five Braginskii viscosity coefficients (with signs chosen to match the stress tensor convention).

The stress tensor is then simply as shown in (\ref{eq:stress_linear_operator})

If we write $\Pi$ as a vector:
\begin{equation}
  \Pi = 
  \begin{bmatrix}
    \Pi_{rr} \\
    \Pi_{r\theta} \\
    \Pi_{rz} \\
    \Pi_{\theta r} \\
    \Pi_{\theta\theta} \\
    \Pi_{\theta z} \\
    \Pi_{zr} \\
    \Pi_{z\theta} \\
    \Pi_{zz}
  \end{bmatrix}
\end{equation}
And if we introduce the index $i$ such that $i = 3(\alpha - 1) + \beta$ then we can map the pair $(\alpha, \beta)$ to a single index to transform the rank 3 tensor $\mathcal{L}_{\alpha\beta j}$ to a rank two tensor $\mathcal{L}_{ij}$ and write:
\begin{equation}
  \Pi_i = -\mathcal{L}_{ij} q_j
  \label{eq:stress_from_P_vector}
\end{equation}

For axisymmetric flows, the divergence of the stress tensor in cylindrical coordinates is:
\begin{equation}
  \nabla \cdot \Pi = 
  \begin{bmatrix}
    \frac{1}{r}\frac{\partial}{\partial r}(r\Pi_{rr}) + \frac{\partial \Pi_{rz}}{\partial z} - \frac{\Pi_{\theta\theta}}{r} \\
    \frac{1}{r}\frac{\partial}{\partial r}(r\Pi_{\theta r}) + \frac{\partial \Pi_{\theta z}}{\partial z} + \frac{\Pi_{r\theta}}{r} \\
    \frac{1}{r}\frac{\partial}{\partial r}(r\Pi_{zr}) + \frac{\partial \Pi_{zz}}{\partial z}
  \end{bmatrix}.
  \label{eq:div_Pi_cylindrical}
\end{equation}

The terms needed in (\ref{eq:div_Pi_cylindrical}) to do the velocity update can now be easily mapped from $\Pi_i$.

\subsection{Coefficient Assembly}

The mapping from the Braginskii formulation to the HYPRE coefficient representation is obtained by expanding the divergence in Eq.~\eqref{eq:div_Pi_cylindrical} and grouping terms according to their differential structure. Each element of $\mathcal{L}_{mj}$ contributes to multiple HYPRE coefficients, depending on how the associated velocity gradients enter the divergence operator.

At each grid point, the magnetic-field unit vector $\mathbf{h} = \mathbf{B}/|\mathbf{B}|$ is evaluated and the viscosity coefficients $\eta_k$ are computed from the local plasma state $(n_i, T_i, \omega_i \tau_i)$. The coefficient tensor $\mathcal{L}_{\alpha\beta j}$ is then constructed using Eqs.~\eqref{eq:Q_tensors} and \eqref{eq:P_construction} by forming the $Q_{k\alpha\beta\mu\nu}$ tensors from the local field geometry and contracting with the transformation matrix $T_{\mu\nu j}$ and the viscosity coefficients.

The HYPRE coefficient tensors $B_{ijkl}$, $E_{ijk}$, $F_{ijk}$, $G_{ij}$, and $W_{ijkl}$ are obtained by expressing the stress tensor in terms of $\mathcal{L}_{mj} q_j$ and systematically expanding its divergence. Terms are grouped according to their differential operator structure and mapped to the corresponding HYPRE coefficients for each velocity component equation.

The resulting sparse linear system is assembled on the staggered grid used in FLASH, with appropriate treatment of boundary conditions, and solved using HYPRE. This approach provides scalable performance for the large systems arising in three-dimensional MagLIF simulations.

\subsection{Energy Equation Coupling}

Following the implicit velocity update, the viscous heating rate \eqref{eq:Qvisc_vector} is computed explicitly using the updated velocity field:
\begin{equation}
  Q_{\text{visc}} = \sum_{k=0}^{2} \eta_k \, W_{k\,ij} \, \nabla_j v_i^{n+1},
  \label{eq:heating_discrete}
\end{equation}
where only the dissipative modes ($k = 0, 1, 2$) contribute to heating. This heating term is then incorporated into the energy equation through operator splitting.

The implicit treatment removes the viscous CFL constraint that would otherwise limit the timestep to
\begin{equation}
  \Delta t_{\text{visc}} \sim \frac{\rho (\Delta x)^2}{\max(\eta_k)},
  \label{eq:visc_cfl}
\end{equation}
which becomes prohibitively restrictive in regions of high viscosity. Instead, the timestep is determined solely by hydrodynamic and magnetic CFL conditions, enabling efficient simulation of the strongly magnetized regime where $\eta_0$ can exceed the perpendicular coefficients by many orders of magnitude.




\section{Verification}
\label{sec:verification}

In this section, we assess the robustness of the implementation through a hierarchy of test problems designed to isolate its key components.

We begin with configurations in which the magnetic field is aligned with the flow, enabling direct comparison to reduced analytical descriptions that retain only the parallel viscosity. We then consider cases where the implementation reduces to known simplified forms of the Braginskii tensor. To quantify convergence, we apply the Method of Manufactured Solutions (MMS) to problems with fully three-dimensional magnetic field topologies, allowing systematic evaluation of spatial and temporal accuracy. Finally, we assess the coupled behavior of the viscosity model within the full hydrodynamic system by comparison with semi-analytic viscous shock solutions.

\subsection{Velocity Diffusion with Magnetic Field Aligned with Velocity Gradient}
\label{sec:firsttest}

We verify our implementation first against the approximate analytical solution derived by Berlok et al. (2020) for magnetized viscosity when the velocity gradients and magnetic field are aligned. This test case provides a good benchmark for the anisotropic momentum diffusion arising from the Braginskii viscosity tensor.

We consider a 2D cylindrical ($r,z$) domain with dimensions $L_z = 0.2$ cm and $L_r = 0.2$ cm, employing axisymmetric boundary condition at axis ($r =0$) and outflow boundary conditions everywhere else. The magnetic field lies in the axial (z) direction, aligned with the velocity gradients. The initial velocity profile is prescribed as a sinusoidal perturbation:
\begin{equation}
v_z(z, t=0) = A \sin(kz),
\label{eq:initial_velocity}
\end{equation}
where $A = 10.0$ cm/s is the amplitude, and the wavenumber is $k = 2\pi/\lambda$ with wavelength $\lambda = L_z$, resulting in a single wavelength fitting within the domain.

For this test, the magnetic field strength is arbitrarily chosen to be a large value of $B = 10^8$ G, and the parallel viscosity coefficient is set to $\eta_0/\rho = 250.0\,\mathrm{cm^2\,s^{-1}}$. While these values are chosen for numerical convenience rather than to match specific MagLIF conditions, they ensure that the viscous diffusion timescale is sufficiently separated from the hydrodynamic timescale to provide a clear test of the viscosity implementation. The simulation employs an adaptive mesh refinement (AMR) structure with a block-based grid covering a domain of $r \in [0.0, L_r]$ and $z \in [0.0, L_z]$. The grid uses square zones ($dr = dz$) with a maximum refinement level of 4. The base resolution is set by $dz = 0.0005$, with coarser blocks refined by a factor of 2 at each level. The domain is divided into blocks containing 8 zones each in both directions.

For this specific geometry where the velocity, its gradients, and magnetic field are all aligned, Berlok et al. (2020) derived an approximate analytical solution. The approximation is due to the fact they use a reduced Braginskii model that only accounts for the parallel viscosity ($\eta_0$). Under these simplifications, the full Braginskii momentum equation reduces to a simpler diffusion equation along field lines, yielding:
\begin{equation}
v_z(z, t) = A \sin(kz) \exp(-\gamma t),
\label{eq:analytical_velocity}
\end{equation}
where the damping rate is:
\begin{equation}
\gamma = \frac{4\eta_{0}}{3\rho}k^2.
\label{eq:damping_rate}
\end{equation}

This solution captures the essential physics of parallel viscous diffusion but neglects several terms present in the full Braginskii formulation: the coupling between different velocity components through off-diagonal stress tensor elements, the viscous heating feedback on the viscosity coefficients through temperature changes, and any magnetic field evolution due to the velocity perturbations. Despite these simplifications, this analytical solution provides a good test case for verifying that our numerical implementation correctly handles the dominant parallel viscosity physics in the aligned configuration.

\begin{figure}[htbp]
    \centering
    \includegraphics[width=0.5\textwidth]{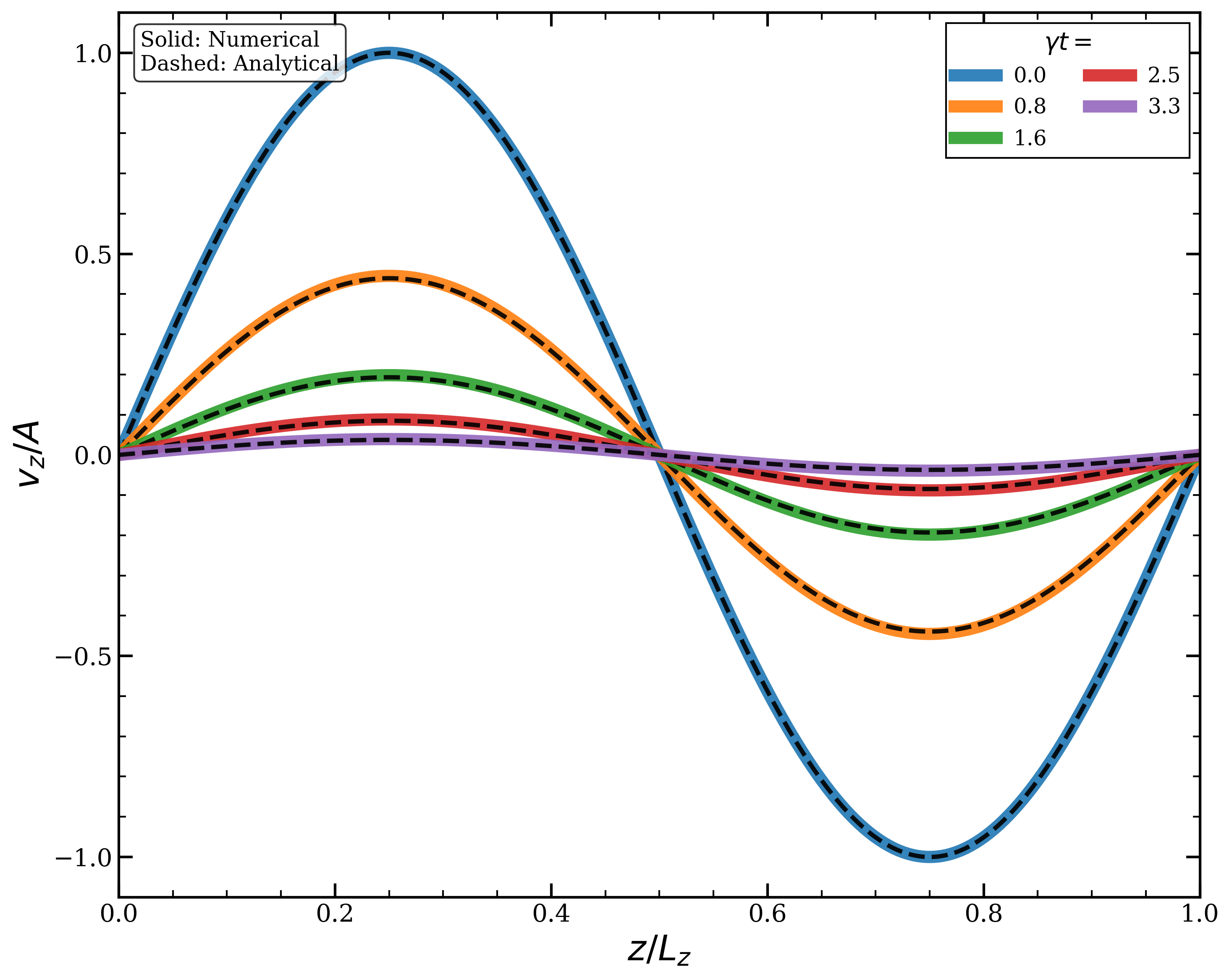}
    \caption{Time evolution of velocity profile showing comparison between FLASH simulation (solid line) and approximate analytical theory (dashed line) at different times when the magnetic field and velocity gradients are parallel to each other.}
    \label{fig:velocity_profile_comparison}
\end{figure}

Figure \ref{fig:velocity_profile_comparison} shows the evolution of the axial velocity profile $v_z(z)$ at several time instances. The numerical solution obtained with our FLASH implementation (solid lines) exhibits good agreement with the analytical prediction from equation \eqref{eq:analytical_velocity} (dashed lines). The velocity amplitude decays exponentially as expected, with the sinusoidal spatial structure preserved throughout the evolution.

\subsection{Verification Against Out-Of-Plane Magnetic Field Implementation}

As an additional verification test, we check whether our implementation reproduces Braginskii's original derivation for the simplified case of a magnetic field aligned solely in the out-of-plane direction. For a magnetic field $\mathbf{h} = (0, 0, 1)$ in Cartesian coordinates, the relevant tensors are:
\begin{equation}
\delta_{\alpha\beta}^{\perp} = \begin{bmatrix}
1 & 0 & 0 \\
0 & 1 & 0 \\
0 & 0 & 0
\end{bmatrix}, \quad
\varepsilon_{\alpha\gamma\beta}h_{\gamma} = \begin{bmatrix}
0 & -1 & 0 \\
1 & 0 & 0 \\
0 & 0 & 0
\end{bmatrix},
\end{equation}
and the tensors $W_{p\alpha\beta}$ take the form:
\begin{equation}
W_{0\alpha\beta} = \begin{bmatrix}
\frac{1}{2}(W_{xx} + W_{yy}) & 0 & 0 \\
0 & \frac{1}{2}(W_{xx} + W_{yy}) & 0 \\
0 & 0 & W_{zz}
\end{bmatrix},
\end{equation}
\begin{align}
W_{1\alpha\beta} &= \begin{bmatrix}
\frac{1}{2}(W_{xx} - W_{yy}) & W_{xy} & 0 \\
W_{yx} & \frac{1}{2}(W_{yy} - W_{xx}) & 0 \\
0 & 0 & 0
\end{bmatrix}, \\
W_{2\alpha\beta} &= \begin{bmatrix}
0 & 0 & W_{xz} \\
0 & 0 & W_{yz} \\
W_{zx} & W_{zy} & 0
\end{bmatrix},
\end{align}
\begin{align}
W_{3\alpha\beta} &= \begin{bmatrix}
-W_{xy} & \frac{1}{2}(W_{xx} - W_{yy}) & 0 \\
\frac{1}{2}(W_{xx} - W_{yy}) & W_{xy} & 0 \\
0 & 0 & 0
\end{bmatrix}, \\
W_{4\alpha\beta} &= \begin{bmatrix}
0 & 0 & -W_{yz} \\
0 & 0 & W_{xz} \\
-W_{zy} & W_{zx} & 0
\end{bmatrix}.
\end{align}
Note that the equations in this section are written in Cartesian coordinates to maintain consistency with Braginskii's original formulation. To implement this test in our cylindrical coordinate system, we align the magnetic field with the $\theta$-direction (out-of-plane) and directly code this simplified analytic form into FLASH. Since the rate-of-strain tensor is defined pointwise at each spatial location, the tensor components themselves retain this form in cylindrical coordinates; the geometric factors associated with the cylindrical frame are automatically incorporated when computing spatial derivatives inside the tensor. It can be shown that by substituting the cylindrical coordinate representation of the magnetic field vector, one recovers the same functional form for the $W_{p\alpha\beta}$ tensors. 

By comparing the results of our general implementation—which handles magnetic fields in arbitrary directions—against this special case, we verify that the code correctly reduces to Braginskii's original expressions when the field is purely azimuthal.

The simulation setup for this test follows the configuration described in Section~\ref{sec:firsttest}, with the following key differences. The domain dimensions are modified to $L_z = 0.1$ cm and $L_r = 0.1$ cm, with the $z$-domain extending from $-L_z$ to $L_z$. Unlike the previous test where the magnetic field was aligned with the velocity gradient, here the magnetic field lies in the azimuthal ($\theta$) direction with a radially varying profile:
\begin{equation}
B_\theta(r) = 10^8r^3 \text{ T},
\label{eq:magnetic_field_profile}
\end{equation}
ensuring the field vanishes at the axis. The initial axial velocity profile is prescribed as:
\begin{equation}
v_z(r, t=0) = 1 - Cr^4 \text{ cm/s},
\label{eq:initial_velocity_radial}
\end{equation}
where $C = 1/L_r^4$, such that the velocity is maximum at the axis and decreases radially outward, vanishing at $r = L_r$.

The simulation employs the same AMR structure with 8 zones per block in each direction and a maximum refinement level of 4, but with a finest resolution of $dz = 0.001$ cm and a minimum refinement level of 3. All other parameters remain unchanged from the previous test.

\begin{figure}[htbp]
    \centering
    \includegraphics[width=0.5\textwidth]{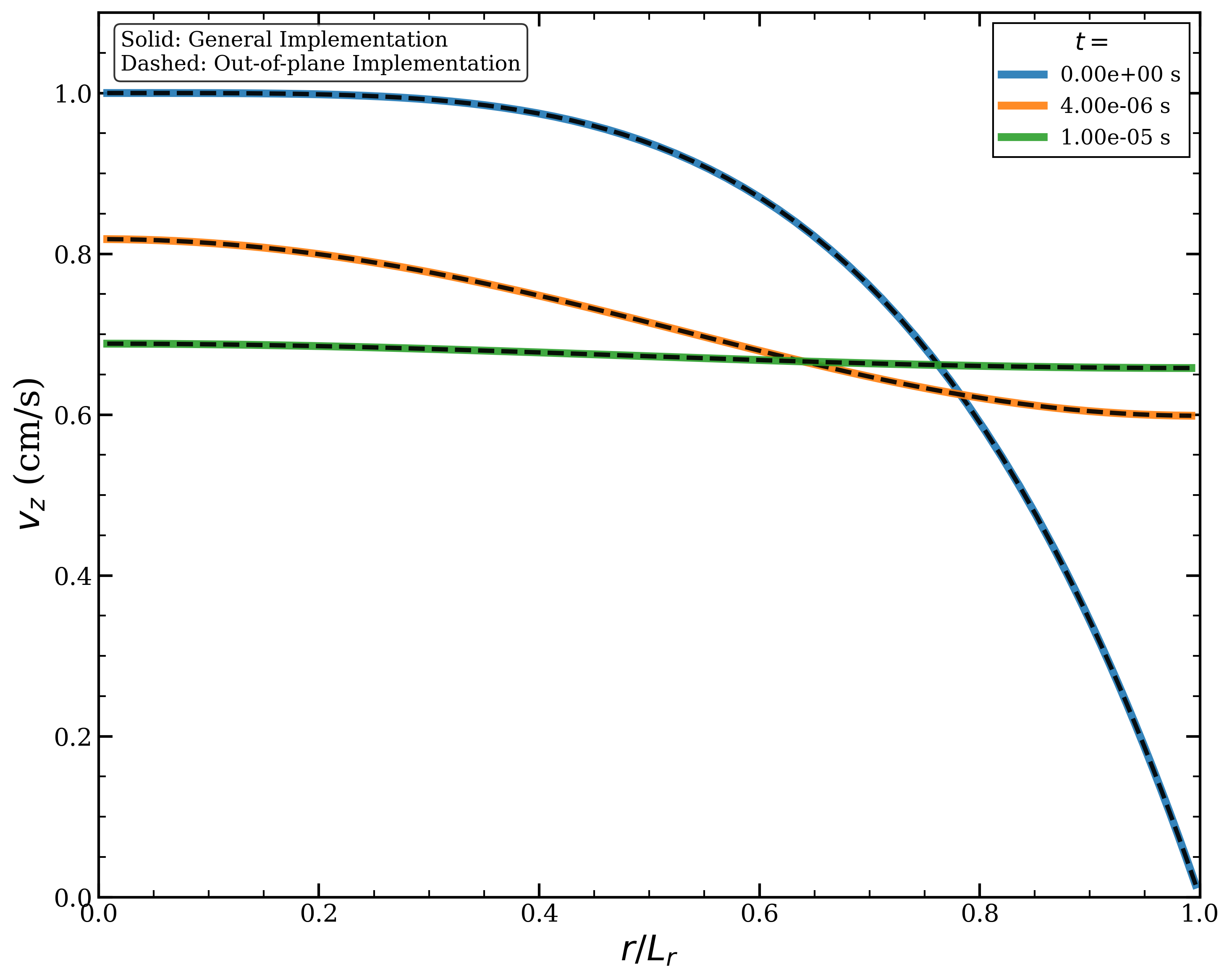}
    \caption{Time evolution of the axial velocity profile comparing our general implementation (solid lines) with the out-of-plane implementation (dashed lines) for an azimuthal magnetic field configuration.}
    \label{fig:azimuthal_field_verification}
\end{figure}

Figure~\ref{fig:azimuthal_field_verification} shows the time evolution of the axial velocity profile, comparing our general implementation (solid lines) against the out-of-plane implementation (dashed lines). The profiles are shown at three different times: the initial condition ($t = 0$ s), an intermediate time ($t = 4 \times 10^{-6}$ s), and a later time ($t = 10^{-5}$ s). The excellent agreement between the two approaches across all time snapshots confirms that our general implementation, which handles arbitrary magnetic field orientations, correctly reduces to Braginskii's original formulation when the field is purely azimuthal. This verification provides confidence that our code is implemented correctly for configurations where the magnetic field lies in the out-of-plane direction.

\subsection{Method of Manufactured Solutions for Diffusion Equation Verification}
\label{sec:mms}

The Method of Manufactured Solutions (MMS) provides a rigorous approach to code verification by constructing problems with known analytical solutions. Unlike validation, which compares simulations against experimental data, verification ensures that the governing equations have been implemented correctly in the code. The MMS approach works by selecting a manufactured solution that is not a true solution to the original governing equations, then computing the residual and setting it as a source term such that the manufactured solution satisfies the modified governing equation exactly. By solving this modified system numerically, we can directly compare the numerical solution against the known manufactured solution and quantify discretization errors.

For our viscosity implementation, we apply MMS to verify the diffusion equation
that governs momentum transport. The governing equation of interest is
equation~(\ref{eq:visc_vel_update}).

We construct a manufactured solution for the velocity field in cylindrical
coordinates:
\begin{equation}
\mathbf{v}^*(r,z,t) = V_0 \, e^{-\gamma t}
\begin{pmatrix}
\sin(k_r \bar{r}) \cos(k_z z) \\
\delta \sin(k_{r\theta} \bar{r}) \sin(k_z z) \\
\cos(k_r \bar{r}) \sin(k_z z)
\end{pmatrix},
\label{eq:manufactured_solution}
\end{equation}
where $\bar{r} \equiv r - r_{\min}$, $k_r \equiv 2\pi/L_r$,
$k_z \equiv 2\pi/L_z$, $k_{r\theta} \equiv 3\pi/(2L_r)$, $\gamma$ is a
temporal decay rate, $V_0 > 0$ is the velocity amplitude, and
$0 < \delta \leq 1$. The three rows correspond to the radial ($v_r$), azimuthal ($v_\theta$), and axial ($v_z$) velocity components, respectively. The exponential time dependence ensures the solution decays smoothly, while the sinusoidal spatial structure provides non-trivial gradients in both the radial and axial directions. 

The manufactured solution is designed to be compatible with the boundary
conditions used in our simulation. At $r = 0$, we impose axisymmetric
boundary conditions, which require $v_r = v_\theta = 0$ and
$\partial v_z / \partial r = 0$. At $r = r_{\max}$, we impose reflecting
boundary conditions, which set $v_r = 0$ and require vanishing normal gradients
of the tangential velocity components. The azimuthal velocity component uses a distinct radial wavenumber $k_{r\theta}$ so that it is not simply proportional to the other components, thereby allowing us to satisfy these boundary conditions. In the axial direction, periodic boundary conditions are applied at both boundaries.

By substituting this manufactured solution into equation~\eqref{eq:visc_vel_update} and computing the full Braginskii stress tensor divergence using all five viscosity coefficients ($\eta_0$ through $\eta_4$), we symbolically determine the required source term $\mathcal{S}$ such that:
\begin{equation}
\rho\frac{\partial\mathbf{v}^*}{\partial t} = -\nabla \cdot
\boldsymbol{\Pi}(\mathbf{v}^*) + \mathcal{S},
\label{eq:mms_with_source}
\end{equation}
where $\mathbf{v}^*$ is the manufactured solution to this modified system. The
source terms were derived symbolically and independently verified.

To establish code verification through MMS, we must demonstrate that the
numerical error systematically decreases as resolution is refined, and that the
observed rate of decrease matches the formal order of accuracy of the
discretization. For a spatial discretization with formal order $p_r$ and a
temporal discretization with formal order $p_t$, the total error is
\begin{equation}
\text{Error} \propto C_r \, \Delta r^{p_r} + C_t \, \Delta t^{p_t},
\end{equation}
where $C_r$ and $C_t$ are constants that depend on the solution but not on the
resolution. By holding $\Delta t$ fixed at a sufficiently small value such that temporal errors are negligible and refining $\Delta r$, the error is dominated by spatial discretization; a log–log plot of error versus $\Delta r$ should therefore exhibit a slope of $p_r$. Conversely, by holding $\Delta r$ fixed at a sufficiently small value such that spatial errors are negligible and refining $\Delta t$, the error is dominated by temporal discretization, and a log–log plot of error versus $\Delta t$ should exhibit a slope of $p_t$. Our spatial discretization is second-order ($p_r = 2$)
and the backward Euler time integrator is first-order ($p_t = 1$).

The error is quantified using the volume-weighted $L_2$ norm appropriate for
cylindrical geometry:
\begin{equation}
\| \mathbf{e} \|_{L_2} = \sqrt{\sum_{i,j} \left( v_{i,j}^{\mathrm{num}}
- v_{i,j}^{*} \right)^2 r_i \, \Delta r \, \Delta z},
\label{eq:l2_error}
\end{equation}
where the sum is over all computational cells $(i,j)$, $v^{\mathrm{num}}$ is the
numerical solution, $v^{*}$ is the manufactured solution evaluated at the same
spatial locations and time, and $r_i$ is the radial coordinate of the cell
center. The factor $r_i \, \Delta r \, \Delta z$ accounts for the cylindrical
volume element, ensuring that cells at larger radii are weighted proportionally
to their physical volume.

We consider a domain with $L_r = L_z = 0.2$~cm and prescribe the magnetic field
with equal components in all three directions:
$\mathbf{B} = (B_r, B_\theta, B_z)$ with $B_r = B_\theta = B_z = 10^8$~G (arbitrarily large). This creates a fully three-dimensional field topology, which couples
all velocity components through the anisotropic stress tensor and exercises the
full implementation including cross-field viscous fluxes. It is important to emphasize that this magnetic field, like the velocity field, is purely manufactured and introduced solely for mathematical verification. It does not satisfy the constraints required for a physically consistent 2D cylindrical geometry. However, our objective here is strictly to verify the implementation, and thus we adopt a field configuration that activates all tensor components without regard to physical realizability. Since these simulations include only the diffusion operator, this choice is mathematically well-posed. The initial velocity is set
according to equation~\eqref{eq:manufactured_solution} at $t = 0$ with
$V_0 = 10$~cm/s, $\delta = 1$, and $\gamma = 6250$~s$^{-1}$. The decay rate
$\gamma$ is chosen so that the solution decays appreciably over the simulation
time while remaining well-resolved throughout the integration, ensuring that the source terms and diffusion operator are exercised over a meaningful dynamic range.

\subsubsection{Spatial Convergence}

To verify the spatial discretization, we hold the timestep fixed at
$\Delta t = 10^{-8}$~s and perform $n_{\mathrm{end}} = 1000$ steps, reaching a
final time of $t_f = 10^{-5}$~s. At this timestep, the temporal error is a fixed
floor that does not vary between runs. We perform five simulations with increasing spatial resolution, using $N_{\mathrm{blocks}} = 1, 2, 4, 8, 16$ blocks per direction (corresponding to $\Delta r = \Delta z = 2.5 \times 10^{-2}$ down to $1.5625 \times 10^{-3}$~cm, with 8 cells per block).

\begin{figure}[htbp]
    \centering
    \includegraphics[width=0.5\textwidth]{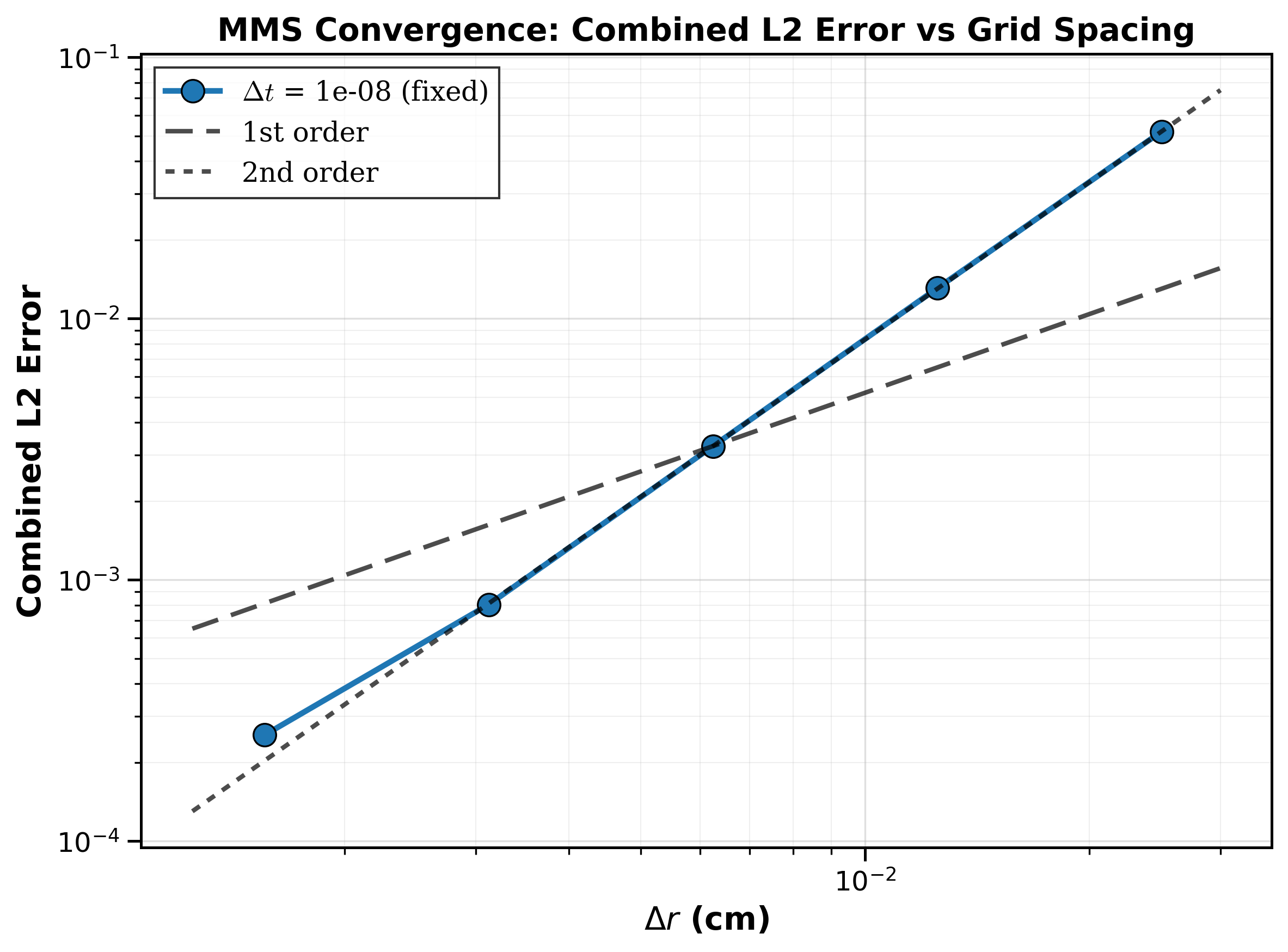}
    \caption{Spatial convergence of the MMS test with all three magnetic field
    components active. The $L_2$ error is plotted against grid spacing $\Delta r$
    for the combined radial, axial, and azimuthal velocity components. The timestep is
    held fixed at $\Delta t = 10^{-8}$~s. Dashed and dotted reference lines
    indicate first- and second-order convergence, respectively. The measured
    rates are approximately 2, confirming second-order spatial accuracy.}
    \label{fig:mms_spatial_convergence}
\end{figure}

Figure~\ref{fig:mms_spatial_convergence} shows the convergence of the $L_2$
error as a function of grid spacing for all three velocity components. The
measured convergence rates between successive refinements are approximately 2.0,
consistent with the expected second-order spatial accuracy of the discretization.
This demonstrates that the spatial operators in the anisotropic viscosity solver
are implemented correctly.

\subsubsection{Temporal Convergence}

To verify the temporal discretization independently, we hold the spatial
resolution fixed at $N_{\mathrm{blocks}} = 32$
($\Delta r = \Delta z = 7.8125 \times 10^{-4}$~cm) and vary the timestep. At
this resolution, the spatial error is a fixed floor. We perform five simulations
with each time step smaller than the next, all integrated to the same final time
$t_f = 6.25 \times 10^{-5}$~s.

\begin{figure}[htbp]
    \centering
    \includegraphics[width=0.5\textwidth]{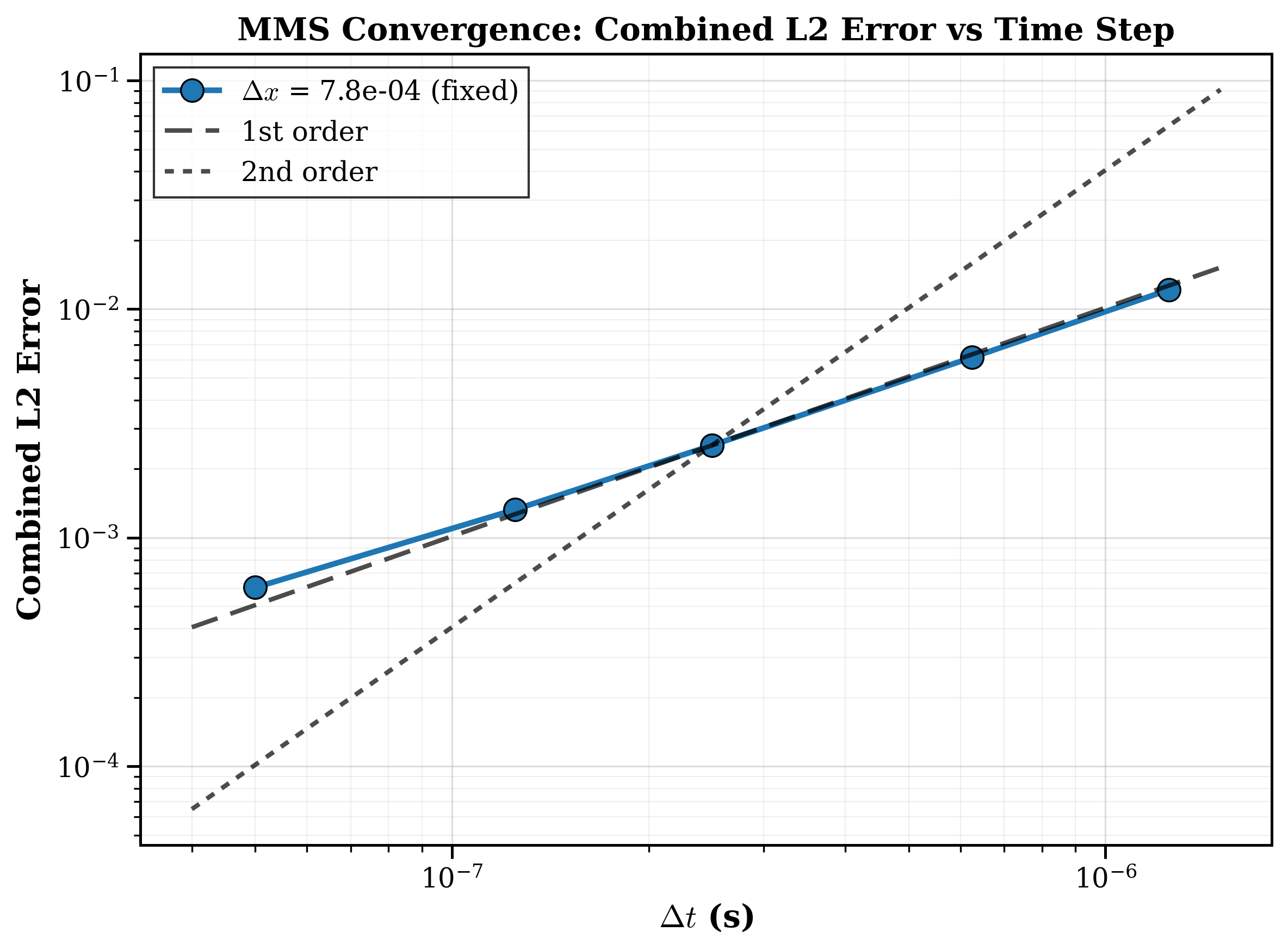}
    \caption{Temporal convergence of the MMS test with all three magnetic field
    components active. The $L_2$ error is plotted against timestep $\Delta t$ for
    the combined radial, axial, and azimuthal velocity components. The spatial resolution
    is held fixed at $\Delta r = 7.8 \times 10^{-4}$~cm. Dashed and dotted
    reference lines indicate first- and second-order convergence, respectively.
    The measured rates are approximately 1, confirming first-order temporal
    accuracy consistent with backward Euler integration.}
    \label{fig:mms_temporal_convergence}
\end{figure}

Figure~\ref{fig:mms_temporal_convergence} shows the convergence of the $L_2$
error as a function of timestep. The measured convergence rates are approximately
1, consistent with the expected first-order accuracy of the backward Euler time
integrator. At the finest timesteps, the error begins to plateau as the spatial
discretization error becomes the dominant contribution.

The MMS verification tests provide quantitative evidence that our Braginskii viscosity implementation correctly solves the anisotropic momentum diffusion
equations. 

\subsection{Magnetized Viscous Shock Profiles}

We finalize the test suite by considering the structure of a collision-dominated shock in a magnetized plasma. 
There are many physical mechanisms shaping the structure of such shocks. 
They include, but are not limited to, electron thermal conductivity, Joule heating, ion and electron viscosities. 
There is extensive literature on various aspects of the theory of plasma shock structures accumulated since the 1950s.\cite{marshall1955structure, zel1957shock, shafranov1957structure, imshennik1962shock, jaffrin1964structure}
We do not intend to review it here. 
Rather, our goal is to construct semi-analytic solutions of the plasma shock structure equations in the case where the only mediating mechanism is viscosity, aimed for MHD codes verification purposes.

We therefore assume that the shock takes place in a fully ionized, singly-charged $Z=1$, quasi-neutral $n_e=n_i=n$, isothermal $T_e=T_i=T$, perfectly conducting hydrogen plasma obeying the ideal-gas equation of state with $\gamma = 5/3$.
Letting $x$ and $u$ be the streamwise direction and velocity, respectively, the only non-zero components of the rate-of-strain tensor in Cartesian geometry are expressed as 

\begin{equation}
    W_{xx} = \frac{4}{3}\frac{\text{d} u}{\text{d} x},\quad W_{yy} = W_{zz} = -\frac{2}{3}\frac{\text{d} u}{\text{d} x}.
    \label{eq:w in shock}
\end{equation}
Assuming that the magnetic field is perpendicular to the direction $x$ of shock propagation, the only relevant component of the viscous stress tensor, $\Pi_{xx}$, is expressed as 

\begin{equation}
    \Pi_{xx} = -\left(\frac{1}{3}\eta_0+\eta_1\right)\frac{\text{d} u}{\text{d} x}.
\end{equation}
We write explicitly the Braginskii expression for the isotropic coefficient, $\eta_0 = 0.96n_iT_i\tau_i$, which is not affected by magnetization.
Introducing the ion Hall parameter $x_i = \omega_i\tau_i$, the $xx$ component of the viscous stress tensor can be invoked as 

\begin{equation}
    \Pi_{xx}=-\frac{4}{3}\eta_0\left[\frac{1}{4}+\frac{3}{4}F\left(x_i\right)\right]\frac{\text{d} u}{\text{d} x},
    \label{eq:pi xx shock}
\end{equation}
where

\begin{equation}
    F\left(x_i\right) \equiv \frac{\eta_1}{\eta_0}=\frac{5x_i^2+2.33}{16x_i^4+16.12x_i^2+2.33}
\end{equation}
is the correction factor for the ion magnetization. 
In the non-magnetized limit, $x_i\ll 1$, there is no anisotropy of viscosity, and the expression in square brackets in \eqref{eq:pi xx shock} tends to unity. 
In the highly-magnetized limit, $x_i\gg1$, $F(x_i)\rightarrow 0$, and the viscous stress term is reduced by a factor of 4.

Under these assumptions, the steady shock structure is governed by the conservation of mass, momentum and energy, and induction, which read

\begin{equation}
    n u = n_1 u_1,
    \label{eq:mass}
\end{equation}

\begin{equation}
\begin{split}
    m_i n u^2 + 2 n T + \frac{B^2}{8\pi} 
    - \frac{4}{3}\eta_0 & \left[ \frac{1}{4} + \frac{3}{4} F(x_i) \right] \frac{\mathrm{d} u}{\mathrm{d} x}\\
    &= m_i n_1 u_1^2 + 2 n_1 T_1 + \frac{B_1^2}{8\pi},
\end{split}
\label{eq:momentum}
\end{equation}

\begin{equation}
\begin{split}
    \frac{1}{2}m_i n u^3 + 5 n T u + \frac{B^2}{4\pi} u
    - & \frac{4}{3}\eta_0 \left[ \frac{1}{4} + \frac{3}{4} F(x_i) \right] u\frac{\mathrm{d} u}{\mathrm{d} x}\\
    &= \frac{1}{2}m_i n_1 u_1^3 + 5 n_1 T_1 u_1 + \frac{B_1^2}{4\pi}u_1,
\end{split}
\label{eq:energy}
\end{equation}

\begin{equation}
    B u = B_1 u_1,
    \label{eq:induction}
\end{equation}
where the subscript $1$ denotes the value of the variables in the far upstream region $x\rightarrow - \infty$. We use these values to derive a viscous length scale characteristic of the pre-shock state as

\begin{equation}
    \Delta_\nu=\frac{\eta_{0_1}}{n_1\sqrt{\frac{10}{3}m_iT_1}}\left[ \frac{1}{4} + \frac{3}{4} F(x_{i_1}) \right],
    \label{eq:viscous length scale}
\end{equation}
which we employ to introduce a dimensionless streamwise coordinate $\xi =x/\Delta_\nu$. 
Accordingly, we normalize the velocity and temperature profiles as $U(\xi)=u/u_1$, $\Theta(\xi) = T/T_1$.
The density and magnetic field profiles can be expressed using mass \eqref{eq:mass} and induction \eqref{eq:induction} as $n/n_1=B/B_1 = 1/U$. 

It is useful to introduce the pre-shock Mach and Alfvén Mach numbers as 
\begin{equation}
  M_1 = \frac{u_1}{\sqrt{10T_1/3m_i}} , \quad M_{A1} = \frac{u_1}{B_1/\sqrt{4\pi m_in_1}},
  \label{eq:machs}
\end{equation}
respectively.
They are related to the pre-shock plasma beta $\beta_1=16\pi n_1T_1/B_1^2$ through $(M_{A_1}/M_1)^2=5\beta_1/6$.
With these definitions, the momentum equation \eqref{eq:momentum} becomes

\begin{equation}
\begin{split}
    U - 1 
    + \frac{3}{5M_1^2}\left(\frac{\Theta}{U} - 1\right)
    &+ \frac{1}{2M_{A_1}^2}\left(\frac{1}{U^2} - 1\right) \\
    &- \frac{4}{3M_1}\mu \Theta^{5/2}
      \frac{\mathrm{d}U}{\mathrm{d}\xi}
    = 0 .
    \label{eq:momentum dimensionless}
\end{split}
\end{equation}
Here, we have neglected the variations of the Coulomb logarithm and assumed $nT\tau \sim T^{5/2}$. 
We have also defined
\begin{equation}
    \mu = \frac{1+3F\left(x_i\right)}{1+3F\left(x_{i_1}\right)}
    \label{eq:mu}
\end{equation}
as the factor accounting for the anisotropy of the viscous stress tensor due to the ion magnetization. 
We can similarly invoke the energy equation \eqref{eq:energy} in a dimensionless fashion as

\begin{equation}
\begin{split}
    \frac{1}{2} \left(U^2 - 1 \right) 
    + \frac{3}{2M_1^2}\left(\Theta - 1\right)
    &+ \frac{1}{M_{A_1}^2}\left(\frac{1}{U} - 1\right) \\
    &- \frac{4}{3M_1}\mu \Theta^{5/2}U
      \frac{\mathrm{d}U}{\mathrm{d}\xi}
    = 0 .
    \label{eq:energy dimensionless}
\end{split}
\end{equation}
From Eqs. \eqref{eq:momentum dimensionless} and \eqref{eq:energy dimensionless}, we express the normalized temperature $\Theta$ via the  normalized velocity $U$ as 

\begin{equation}
    \Theta\left(U\right) = 1+\frac{1}{9}\left(U-1\right)\left[5M_1^2\left(U-1\right)-6\right] - \frac{5M_1^2\left(U-1\right)^2}{9M_{A_1}^2U}.
    \label{eq:Theta via U}
\end{equation}
Substituting \eqref{eq:Theta via U} into \eqref{eq:energy dimensionless}, we finally obtain the equation governing the structure of a magnetized shock mediated by viscosity:

\begin{equation}
\begin{split}
    &\frac{1}{M_1}\Theta(U)^{5/2}
    \mu U \frac{\mathrm{d}U}{\mathrm{d}\xi} \\
    &= \frac{U-1}{U}\left[
        U^2
        - \left(\frac{1}{4}
        + \frac{3}{4M_1^2}
        + \frac{5}{8M_{A_1}^2}\right) U
        - \frac{1}{8M_{A_1}^2}
    \right].
\end{split}
\label{eq:shock structure}
\end{equation}

To finalize the derivation, we need to express the anisotropy factor $\mu$ via the normalized velocity $U$. 
Noting that $x_i \sim B T^{3/2}/n$, and that the ratio $B/n$ remains constant across the shock structure, the ion Hall parameter can be expressed in terms of its value upstream $x_{i_1}$ and the normalized velocity $U$ as $x_i = x_{i_1}\Theta\left(U\right)^{3/2}$.
In other words, the ion Hall factor in the shock front does not increase because of magnetic flux compression, since the plasma density increases proportionally to the magnetic field. 
Increase in the ion magnetization occurs entirely because of the ion temperature rise. 

The right-hand side of Eq. \eqref{eq:shock structure} vanishes for $U=1$, which corresponds to the pre-shock state. The post-shock state $U_2$ is given by the only other positive root of its right-hand side:

\begin{equation}
\begin{split}
    U_2 &= \frac{M_1^2+3}{8M_1^2} + \frac{5}{16M_{A_1}^2} \\
        &\quad + \left[ 
            \left(\frac{M_1^2+3}{8M_1^2} + \frac{5}{16 M_{A_1}^2}\right)^2
            + \frac{1}{8M_{A_1}^2} 
        \right]^{1/2} .
\end{split}
\label{eq:U2}
\end{equation}
Substituting $U=U_2$ into \eqref{eq:Theta via U} yields the post-shock normalized temperature $\Theta_2$. 
Notice that the pre-shock ion Hall parameter $x_{i_1}$ does not influence the normalized post-shock variables. 
Rather, it only modifies the shock structure through the anisotropy factor $\mu$. 
The shock structure is obtained by numerical integration of Eq. \eqref{eq:shock structure}.
We choose to place the origin $\xi=0$ at the average value of the normalized velocity, $U=(1+U_2)/2$.

The effect of the viscosity anisotropy on the shock structure can therefore be significant if (a) the shock is strong enough, so that the isotropic viscosity increase in the shock front is substantial, and, (b) the pre-shock ion Hall parameter is moderately weak so that the factor $\mu$ is allowed to vary within the shock structure. 
Effectively, if viscosity were anisotropic everywhere, the resulting shock structure would be identical to that of an unmagnetized shock, with the only difference residing in the viscous length scale $\Delta_\nu$.

To generate a numerical example, we choose the parameters of a strong transverse MHD shock $M_1 = 5$ and $\beta_1=4$ or, equivalently, $M_{A_1} = 9.129$. 
This results in a shock density compression and temperature jump of $1/U_2 = 3.42$ and $\Theta_2=8.14$, respectively.
We compare two cases of pre-shock weakly magnetized ions: $x_{i_1}= 0.01$ and $x_{i_1} = 0.2$. 
In the former case, the post-shock plasma is weakly magnetized with $x_{i_2}=0.232$, and the viscosity anisotropy parameter remains close to unity $\mu=0.851$.
In the latter case, the post-shock plasma is magnetized: $x_{i_2}=4.64$, and the corresponding viscosity anisotropy parameter is noticeably less than unity $\mu_2=0.295$.

\begin{figure}[htbp]
    \centering
    \includegraphics[width=0.95\linewidth]{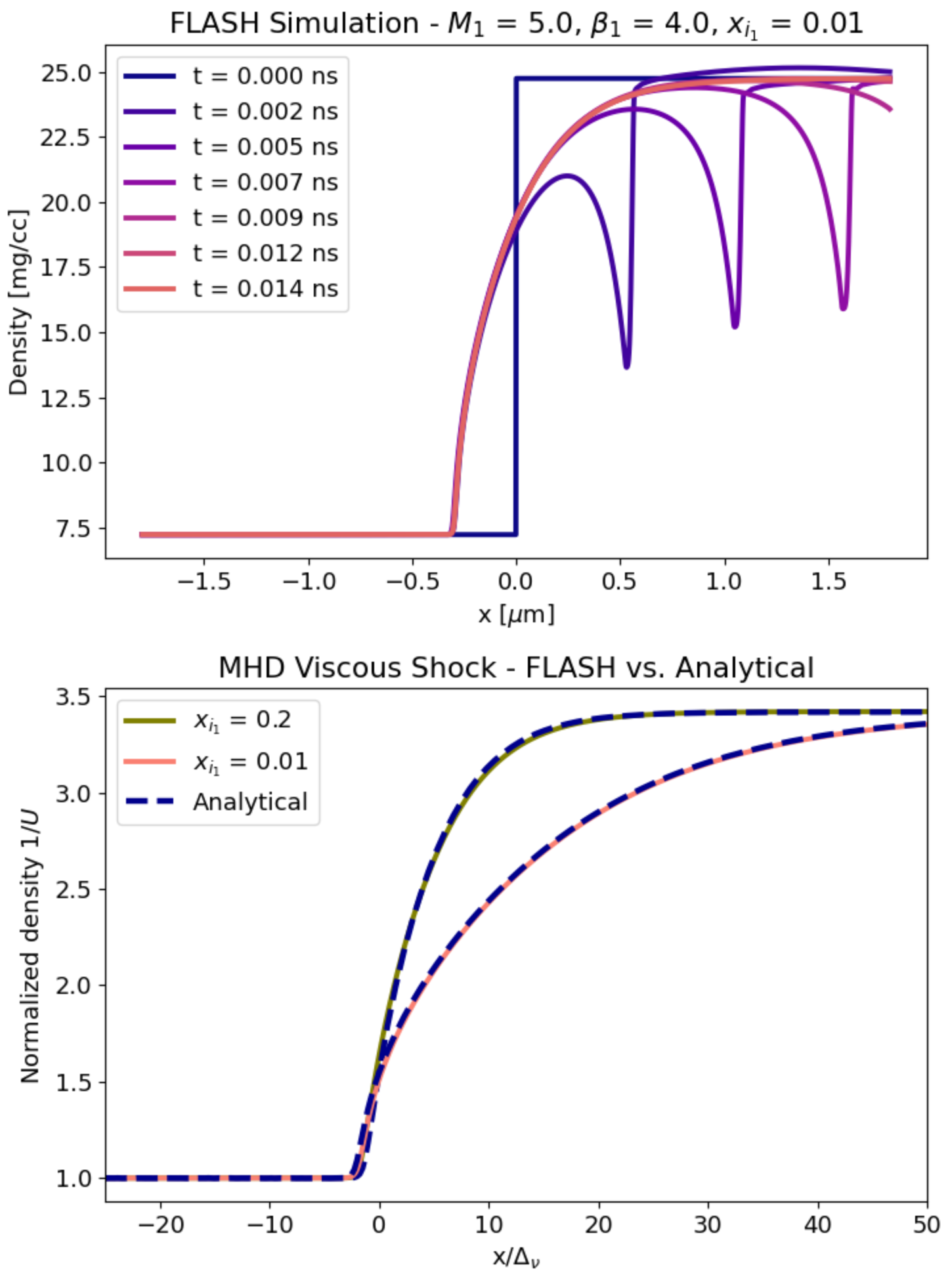}
    \caption{FLASH simulation of MHD viscous shock formation from a Heaviside-step initial condition. Top panel: temporal snapshots showing the evolution until the shock profile is fully developed. Bottom panel: comparison of the fully developed shock profile with analytical solutions.}
    \label{fig:shock verification}
\end{figure}

FLASH simulations of shock formation for these parameter choices are shown in Fig. \ref{fig:shock verification}. 
These simulations were initialized with a Heaviside-step initial condition, separating two fluid states related by the Rankine-Hugoniot jump conditions.
A pre-shock pressure level $p_1$ of 1 Mbar has been chosen arbitrarily. 
The remaining quantities are dictated by the choice of Mach, beta, and ion Hall parameters, and are specified in Table \ref{tab:viscous_shock_inputs}.
We have chosen a constant Coulomb logarithm value of 7.0.
The simulation setup considers cylindrical geometry with the Heaviside function placed far from the origin for the geometric factors to not play any role. 
The top panel in Fig. \ref{fig:shock verification} displays the shock formation process for the $x_{i_1} = 0.01$ case dictated by viscous diffusion. 
It can be seen that the shock profile remains stationary after full formation. 
The bottom panel compares the fully established shock profiles for both magnetization cases with the solution of the analytical model [Eqs. \eqref{eq:Theta via U} and \eqref{eq:shock structure}], demonstrating excellent agreement. 
It can be seen how the shock width is noticeably lower in the $x_{i_1} = 0.2$ case because of the reduction of the viscosity due to magnetization.

\begin{table*}[t]
\centering
\begin{tabular}{c||cccc|cccc}
\hline
 & \multicolumn{4}{c|}{Pre-shock} & \multicolumn{4}{c}{Post-shock} \\
Case $x_{i_1}$ 
& $\rho_1$ [mg/cm$^3$] & $p_1$ [Mbar] & $u_1$ [km/s] & $B_1$ [MG]
& $\rho_2$ [mg/cm$^3$] & $p_2$ [Mbar] & $u_2$ [km/s] & $B_2$ [MG] \\
\hline\hline
0.01 
& 7.24 & 1.00 & 758.67 & 2.51
& 24.74 & 27.80 & 222.01 & 8.57 \\
\hline
0.2 
& 2.18 & 1.00 & 1381.20 & 2.51
& 7.46 & 27.80 & 404.17 & 8.57 \\
\hline
\end{tabular}
\caption{Pre- and post-shock MHD quantities used for viscous shock verification.}
\label{tab:viscous_shock_inputs}
\end{table*}

\section{Impact of Magnetized Viscosity}
\label{sec:impact}

In this section, we investigate the effect of magnetized viscosity on MagLIF simulations. We present two representative MagLIF configurations, each simulated with and without the magnetized viscosity module enabled, to evaluate the influence of anisotropic viscous transport on implosion dynamics and fusion performance.

\subsection{Pool Heated MagLIF Simulation}
\label{sec:pool_heated}

The first simulation represents an example of a MagLIF target driven by the Pacific Fusion Demonstration System.\cite{alexander_affordable_2025} In this configuration, preheat energy is deposited into a plug of DT ice at the base of the target, launching hot, magnetized plasma upward to fill the target volume during the implosion phase.\cite{Sefkow2014MagLIF} The complex flow structures that emerge in this scenario can drive nonuniformities within the target that degrade performance. We hypothesize that plasma viscosity plays a significant role in smoothing and dissipating these structures, potentially improving simulated target performance substantially.

\begin{figure}[htbp]
    \centering
    \includegraphics[width=0.5\textwidth]{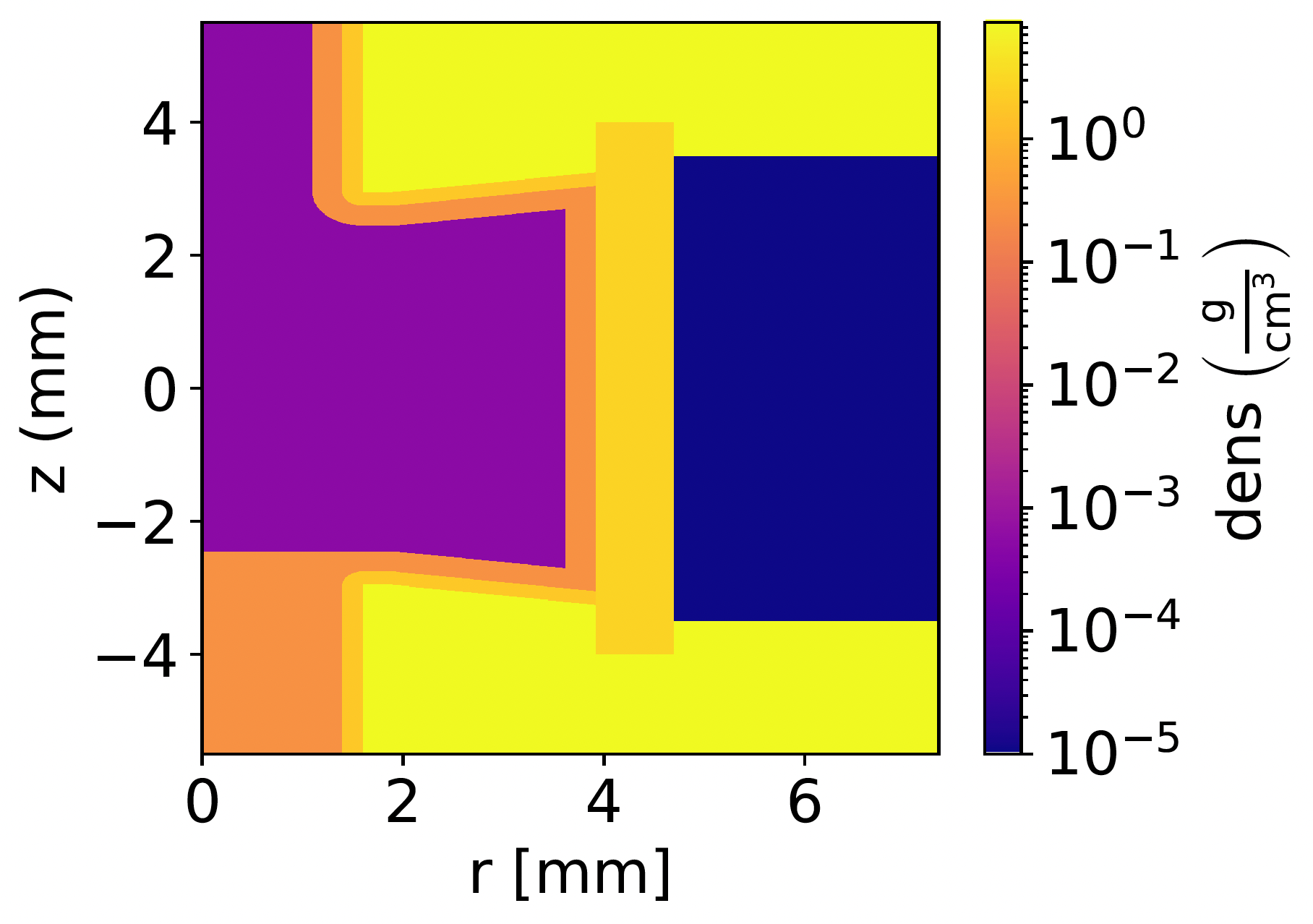}
    \caption{Initial target configuration for the pool heated MagLIF simulation showing the density profile. The aluminum liner (yellow, $\sim$2.7 g/cc) surrounds a DT ice layer (orange, 0.25 g/cc) coating the inner liner wall, with a central DT vapor fill (magenta, $\sim$1 mg/cc). The dark blue region represents the vacuum ($\sim$10$^{-5}$ g/cc) surrounding the target. An initial axial magnetic field of 10 T is applied throughout the domain.}
    \label{fig:initial_config}
\end{figure}

We configured a 2D axisymmetric simulation of this system with an in-plane magnetic field, as shown in Figure~\ref{fig:initial_config}. The target consists of an aluminum liner (initial density 2.7 g/cc) surrounding a DT ice layer (density 0.25 g/cc) that coats the inner liner surface, with a central DT vapor region (initial density $\sim$1 mg/cc). An initial axial magnetic field of 10 T provides magnetic insulation of the fuel. Preheat energy is deposited using an idealized model that distributes energy within a cylindrical region of radius 843 $\mu$m, extending from the laser entrance hole (LEH) at the top of the target to halfway into the ice plug; the deposition is weighted by zonal mass, which biases most of the preheat energy into the denser ice layer rather than the vapor. For this study, we use an elevated preheat energy of 100 kJ/cm—approximately three times a representative preheat level used in MagLIF-scale simulations, consistent with scaling studies at higher drive currents \cite{ruiz_similarity_2023}—to accentuate the differences between viscous and inviscid simulations. The simulation was driven to a peak current of 58 MA with alpha particle deposition disabled, so the simulation does not reach ignition and no fusion yield is produced; this configuration isolates the hydrodynamic effects of viscosity from burn physics. The simulation was executed on 192 CPUs, with and without the viscosity module enabled.

\begin{figure*}[htbp]
    \centering
    \includegraphics[width=0.48\textwidth]{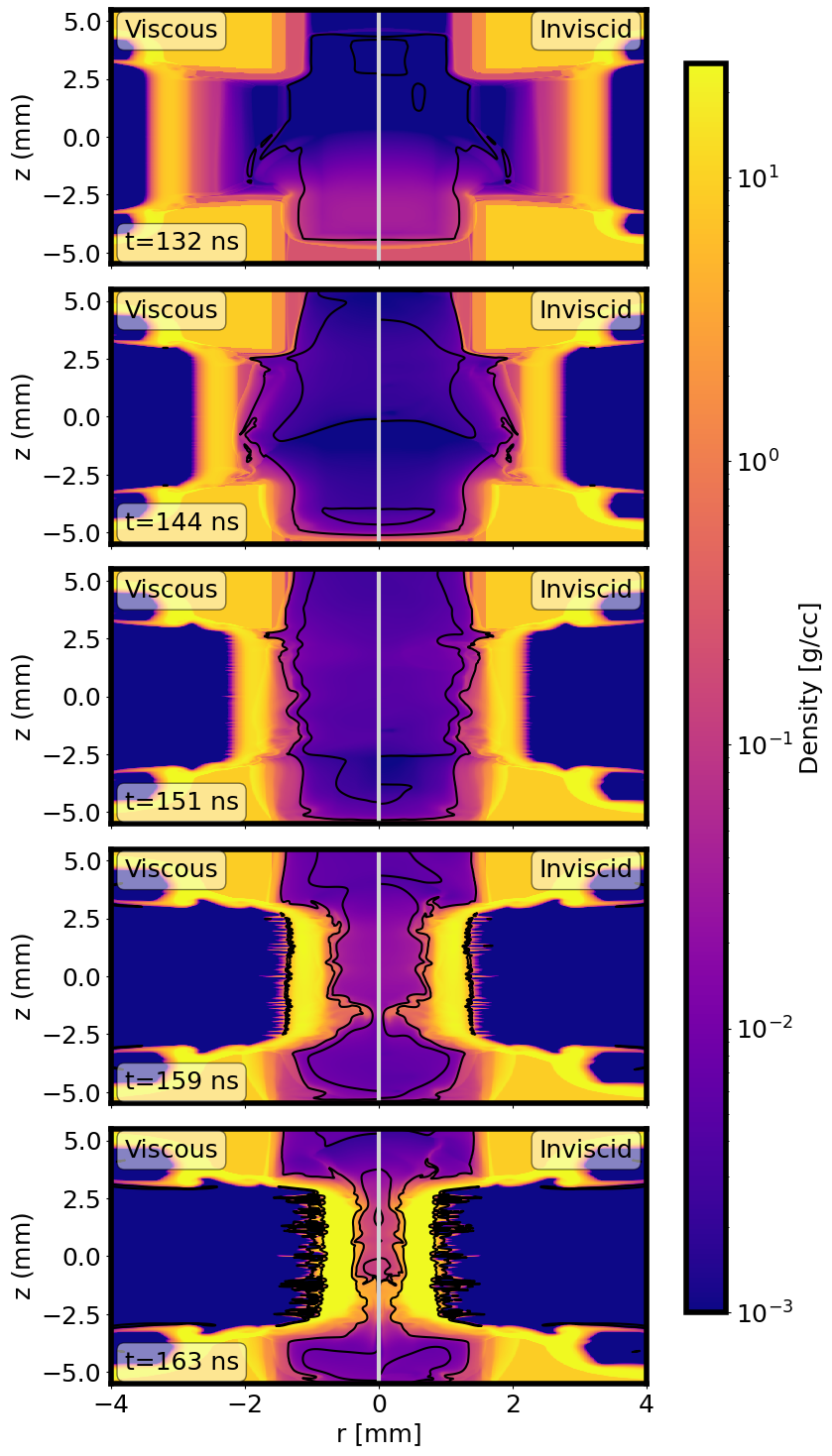}
    \hfill
    \includegraphics[width=0.48\textwidth]{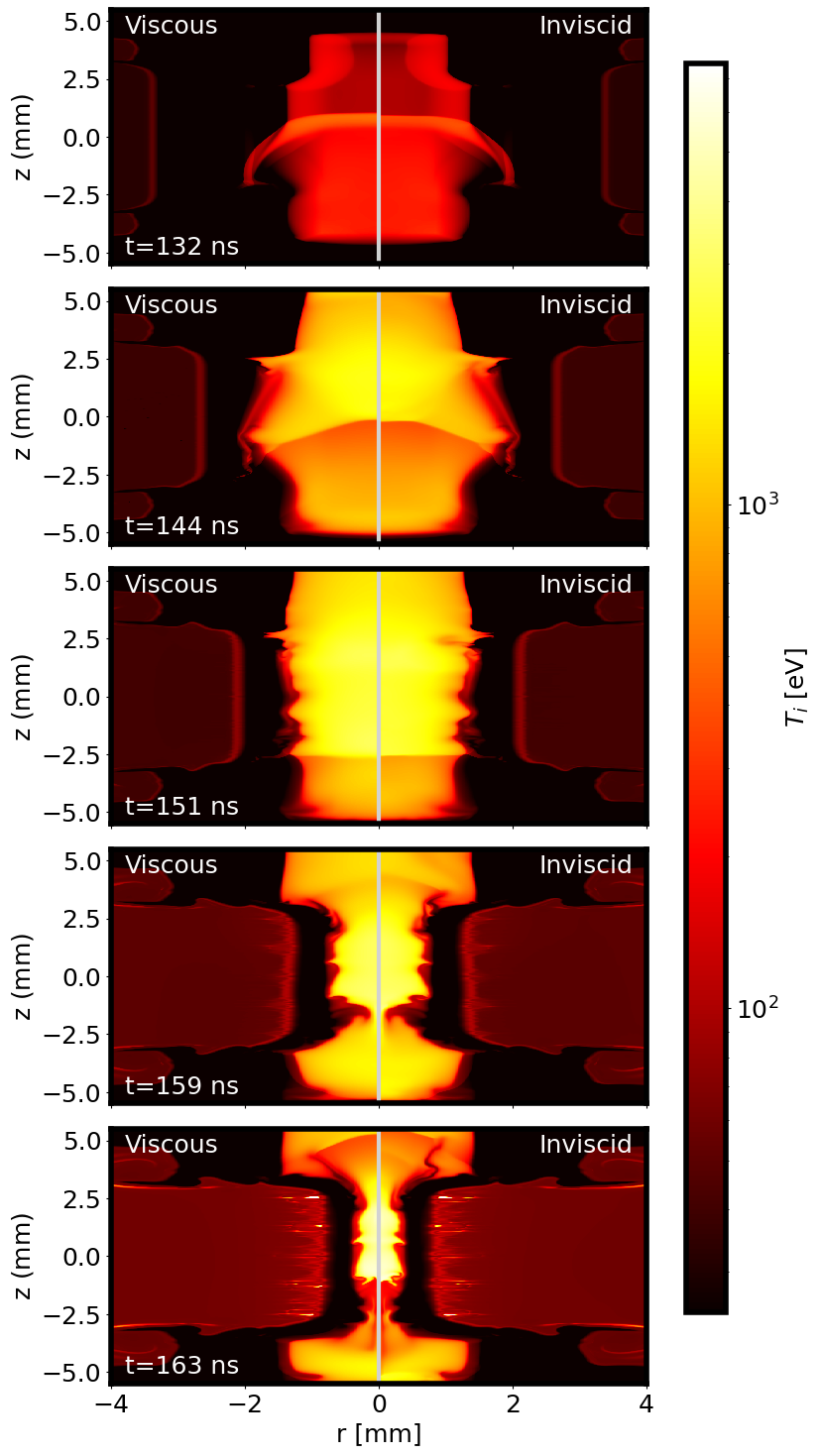}
    \caption{Time evolution of the density field (left) and ion temperature field (right) for the pool heated MagLIF simulation, comparing viscous (left half of each panel) and inviscid (right half) simulations. The sequence spans from $t = 132$~ns through stagnation at $t = 163$~ns. The black isoline on the density plots marks the $T_i = 1$~keV contour, overlaid to delineate fine-scale thermal structures in the hot spot region where density contrast is minimal.}
    \label{fig:pf_demo_evolution}
\end{figure*}

Figure~\ref{fig:pf_demo_evolution} presents the temporal evolution of the density and ion temperature fields for both the viscous and inviscid simulations. To facilitate direct comparison, the viscous simulation results have been mirrored about the axis and placed adjacent to the inviscid results, with the viscous case shown on the left half and the inviscid case on the right half of each panel. These evolutions reveal several notable differences between the two cases. Close inspection of the temperature plots shows that the upward-propagating shock front is slightly thicker in the viscous simulation (at $t = 132$~ns), consistent with the well-established result that viscosity increases the characteristic width of shock structures~\cite{ZeldovichRaizer_Shock}. At $t = 163$~ns, wispy filamentary structures appear in the inviscid simulation near $r \approx 1$~mm, $z \approx 4$~mm that are notably absent in the viscous case, indicating that viscosity actively smooths fine-scale flow features. To delineate differences in fine-scale thermal structures that are difficult to discern from the density field alone---particularly in the hot spot where there is little density contrast---we overlay a 1~keV ion temperature isoline on the density plots. 

To provide a more detailed comparison, we examine the vorticity field, defined as
\begin{equation}
    \boldsymbol{\omega} = \nabla \times \mathbf{v},
    \label{eq:vorticity}
\end{equation}
where $\mathbf{v}$ is the velocity field. In our 2D axisymmetric geometry, the relevant component is the azimuthal vorticity $\omega_\theta = \partial_z v_r - \partial_r v_z$.

\begin{figure}[htbp]
    \centering
    \includegraphics[width=\columnwidth]{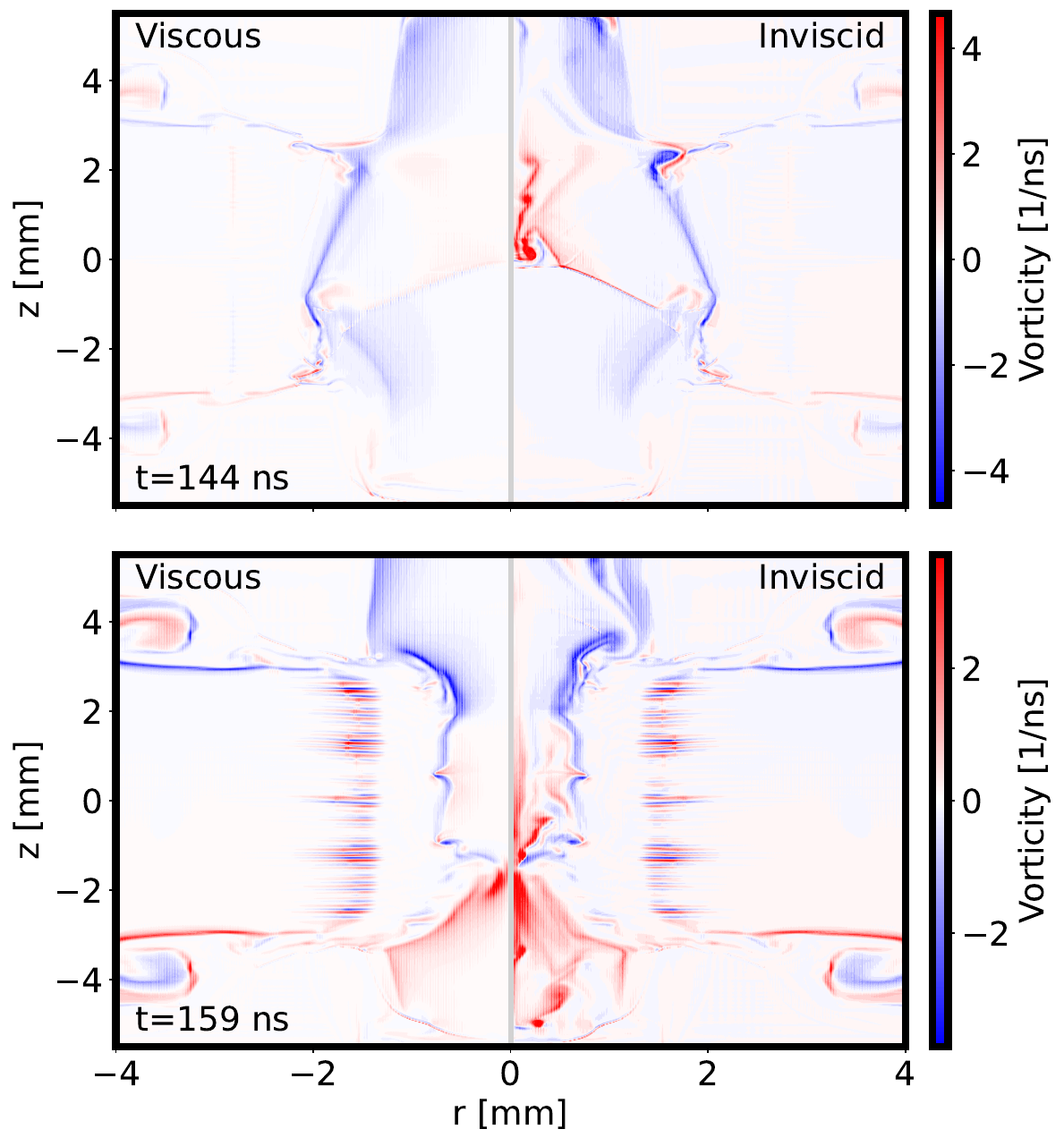}
    \caption{Vorticity field comparison between viscous (left half) and inviscid (right half) simulations at $t = 144$~ns and $t = 159$~ns. The inviscid simulation exhibits substantially higher net vorticity throughout the domain, particularly in regions of strong shear flow. The viscous simulation demonstrates effective damping of these rotational structures, consistent with the expected dissipative action of the Braginskii viscosity tensor.}
    \label{fig:vorticity}
\end{figure}

Figure~\ref{fig:vorticity} presents a comparison of the vorticity field between the viscous and inviscid simulations at two representative times. The inviscid simulation exhibits substantially higher net vorticity throughout the domain, with prominent rotational structures developing in regions of strong velocity shear. In contrast, the viscous simulation demonstrates markedly reduced vorticity, confirming that magnetized viscosity effectively damps these rotational flow structures. This result is consistent with our initial hypothesis that viscosity can suppress the formation and persistence of vortical features that would otherwise develop in inviscid simulations.

\begin{figure*}[htbp]
    \centering
    \includegraphics[width=\textwidth]{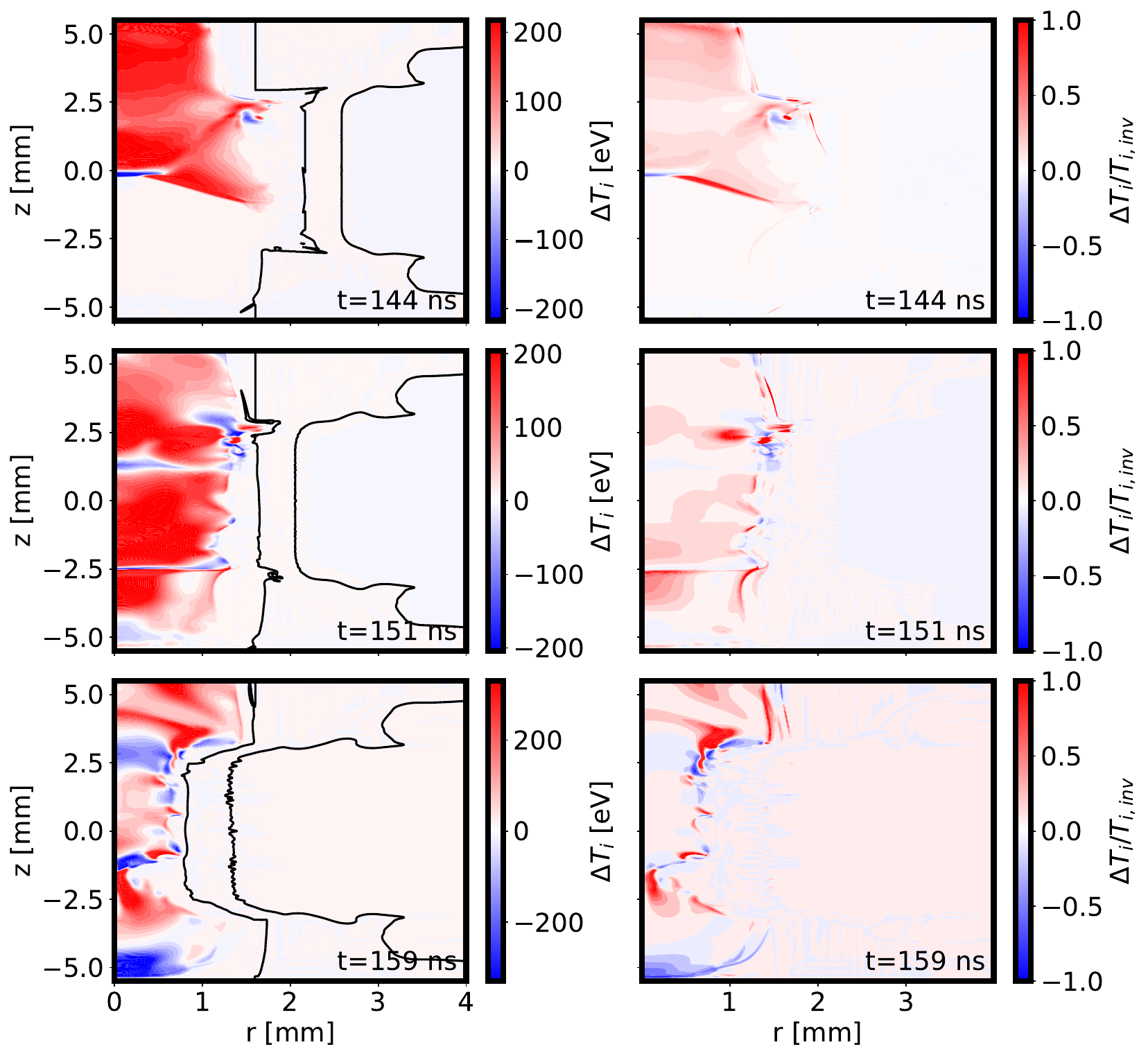}
    \caption{Ion temperature differences between viscous and inviscid simulations at three times during the implosion. Left column: absolute temperature difference $\Delta T_i = T_i^{\text{viscous}} - T_i^{\text{inviscid}}$ in eV. Right column: relative temperature difference $\Delta T_i / T_{i,\text{inv}}$ normalized by the local inviscid temperature. Red regions indicate where the viscous simulation is hotter; blue regions indicate where it is cooler. The black contour lines in the left column denote the aluminum liner boundary. Temperature enhancements of order 100--200 eV (corresponding to relative increases of 50--100\% in some regions) are visible in the fuel, particularly near the liner interface and in regions of strong velocity shear.}
    \label{fig:temp_diff}
\end{figure*}

Beyond merely damping vortical structures, viscosity converts the kinetic energy stored in these rotational motions into thermal energy of the plasma. This energy conversion pathway has important implications for the thermal state of the fuel. Figure~\ref{fig:temp_diff} presents the ion temperature difference between the viscous and inviscid simulations, defined as $\Delta T_i = T_i^{\text{viscous}} - T_i^{\text{inviscid}}$.

The temperature difference shown in Figure~\ref{fig:temp_diff} reveals that the viscous simulation achieves consistently higher temperatures throughout much of the plasma volume. The left column displays the absolute temperature difference $\Delta T_i = T_i^{\text{viscous}} - T_i^{\text{inviscid}}$, while the right column shows the relative difference normalized by the local inviscid temperature, $\Delta T_i / T_{i,\text{inv}}$. This normalization provides a clearer picture of how viscosity alters the temperature evolution: while absolute differences of 100--200 eV may appear modest, they represent relative enhancements of 50--100\% in cooler regions of the fuel near the liner interface. We attribute this temperature increase to two complementary mechanisms: direct viscous heating through the $Q_{\text{visc}} = -\boldsymbol{\Pi} : \nabla \mathbf{v}$ term in the energy equation, and the conversion of kinetic energy stored in vortical structures into thermal energy as these structures are damped by viscous dissipation. Both mechanisms represent favorable energy conversion pathways that redirect energy that would otherwise be lost in unproductive fluid motions into useful thermal energy that can contribute to fusion reactions.

These results demonstrate that magnetized viscosity has a non-negligible and beneficial effect on MagLIF simulations. The suppression of vortical structures and the associated thermal energy enhancement represent favorable outcomes for fusion performance. However, the computational expense of this simulation precludes extensive parameter sweeps to systematically study the effect of viscosity on perturbation growth and yield. Therefore, to conduct yield sensitivity studies, we employ a computationally more tractable configuration described in the following subsection.

\subsection{Traditional MagLIF Configuration with Seeded Perturbations}

To systematically investigate the impact of magnetized viscosity on instability growth and fusion yield, we employ a simplified MagLIF configuration that is computationally less demanding. This simulation represents a traditional MagLIF setup with the domain initially partitioned (proceeding radially outward) into: DT gas, DT ice, an aluminum liner, and vacuum. Magnetic pressure drives the liner inward, compressing the fuel to achieve fusion conditions.

We configured a 2D axisymmetric MagLIF implosion with an in-plane magnetic field. The target consists of a central low-density DT gas column surrounded by a DT ice annulus at 0.25~g/cc, which is in turn enclosed by an aluminum liner at 2.7~g/cc; material outside the liner is treated as low-density vacuum. An initially uniform 15~T axial magnetic field provides magnetic insulation of the fuel. Fuel preheat is modeled as a volumetric energy source that deposits 30~kJ/cm between 115 and 123~ns within a cylindrical region of radius 843~$\mu\mathrm{m}$, uniformly over the axial extent of the periodic wedge; although the deposition is formally weighted by zonal mass, the nearly uniform density in this configuration makes the preheat effectively uniform throughout the preheated volume. To seed magneto–Rayleigh–Taylor instability, we impose a single-mode sinusoidal perturbation with wavelength 200~$\mu\mathrm{m}$ on the inner surface of the ice layer. The implosion is driven to a peak current of approximately 60~MA and evolved through peak compression with thermonuclear burn and alpha-particle energy deposition enabled; unlike the pool-heated configuration in Section~\ref{sec:pool_heated}, these standard MagLIF simulations include self-consistent alpha deposition, and we perform paired runs with and without the magnetized viscosity model to assess its influence on instability growth and overall fusion performance.

Such implosions are susceptible to magneto-Rayleigh-Taylor (MRT) instabilities arising from the acceleration of the dense liner by the low-density magnetic field region, followed by the deceleration of the dense ice and liner material as pressure peaks in the low-density hot spot at stagnation. This latter deceleration-phase instability is where viscosity may play its most significant role. Classical theory predicts that viscosity reduces the growth rate of these instabilities by damping small-scale velocity perturbations~\cite{chandrasekhar_hydrodynamic_2013}. To test this hypothesis in the MagLIF context, we seed these instabilities with sinusoidal perturbations on the inner surface of the ice layer. We conduct a series of simulations with varying initial perturbation amplitudes (0, 10, 20, 30, and 40~$\mu$m) at a fixed wavelength of 200~$\mu$m, running each case with and without magnetized viscosity enabled. We note that these amplitudes far exceed the anticipated roughness for a real DT ice layer and might be considered an ``overtest'' for assessing the impact of viscous damping on MRT feedthrough into the hot spot. Additionally, the outer liner surface is initialized as smooth with no perturbations seeded in density or temperature; consequently, any instabilities that emerge on the liner outer surface are seeded by flow asymmetries produced by the outward-propagating preheat blast wave rather than by imposed surface roughness.

\begin{figure*}[t]
    \centering
    \includegraphics[width=\columnwidth]{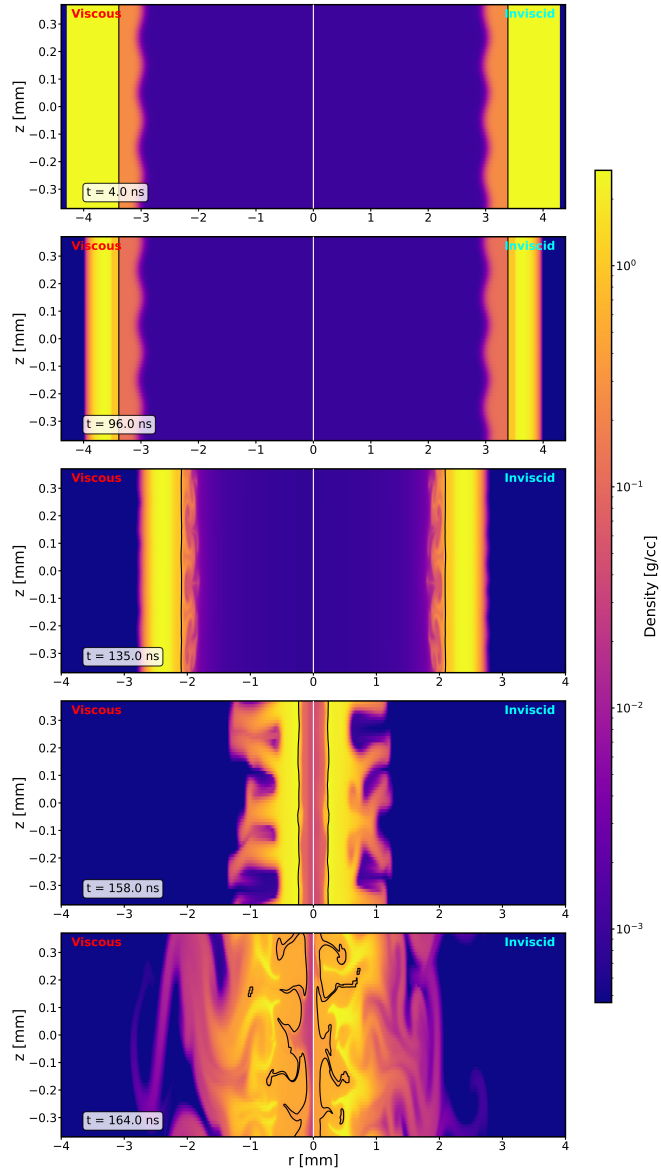}
    \caption{Temporal evolution of the density field comparing viscous (left half of each panel) and inviscid (right half) simulations for an initial perturbation amplitude of 40~$\mu$m. The sequence spans from early conditions ($t = 4$~ns) through the onset of the explosion phase ($t = 164$~ns). The most significant difference emerges at late times near the axis ($r = 0$): the viscous simulation maintains a coherent low-density hot spot, while in the inviscid case the hot spot has essentially collapsed and filled with higher-density material due to hydrodynamic mixing from deceleration-phase MRT instabilities. Secondary differences include the suppression of vortical structures near $r = 0.6$~mm, $z = -0.1$~mm at $t = 158$~ns in the viscous case, and reduced mode coupling among the trailing liner MRT spikes at $t = 164$~ns. The black contours represent the aluminum boundary.}
    \label{fig:dens_evolution}
\end{figure*}

Figure~\ref{fig:dens_evolution} presents the density field evolution for a representative case with an initial perturbation amplitude of 40~$\mu$m. The sequence spans from early conditions at $t = 4$~ns through the beginning of the explosion phase at $t = 164$~ns. During the early stages of the implosion ($t = 4$, 96, and 135~ns), no appreciable visual differences appear between the viscous and inviscid simulations, as the instabilities have not yet grown to amplitudes where viscous damping becomes significant. However, at later times ($t = 158$ and 164~ns), dramatic differences emerge. Most critically, the viscous simulation maintains a coherent low-density hot spot near the axis ($r = 0$) through the final time shown, whereas the inviscid simulation shows that this volume has completely filled with higher-density material---the hot spot has essentially collapsed. This collapse is presumably due to catastrophic hydrodynamic mixing induced by deceleration-phase MRT instabilities that are not damped in the absence of viscosity. The trailing MRT spikes of liner material, while visually prominent, are dynamically decoupled from the critical stagnating fuel assembly at bang time; nevertheless, the viscous simulation also exhibits reduced mode coupling among these structures, with individual RT fingers remaining more distinct compared to the merged structures in the inviscid case.

\begin{figure}[htbp]
    \centering
    \includegraphics[width=\columnwidth]{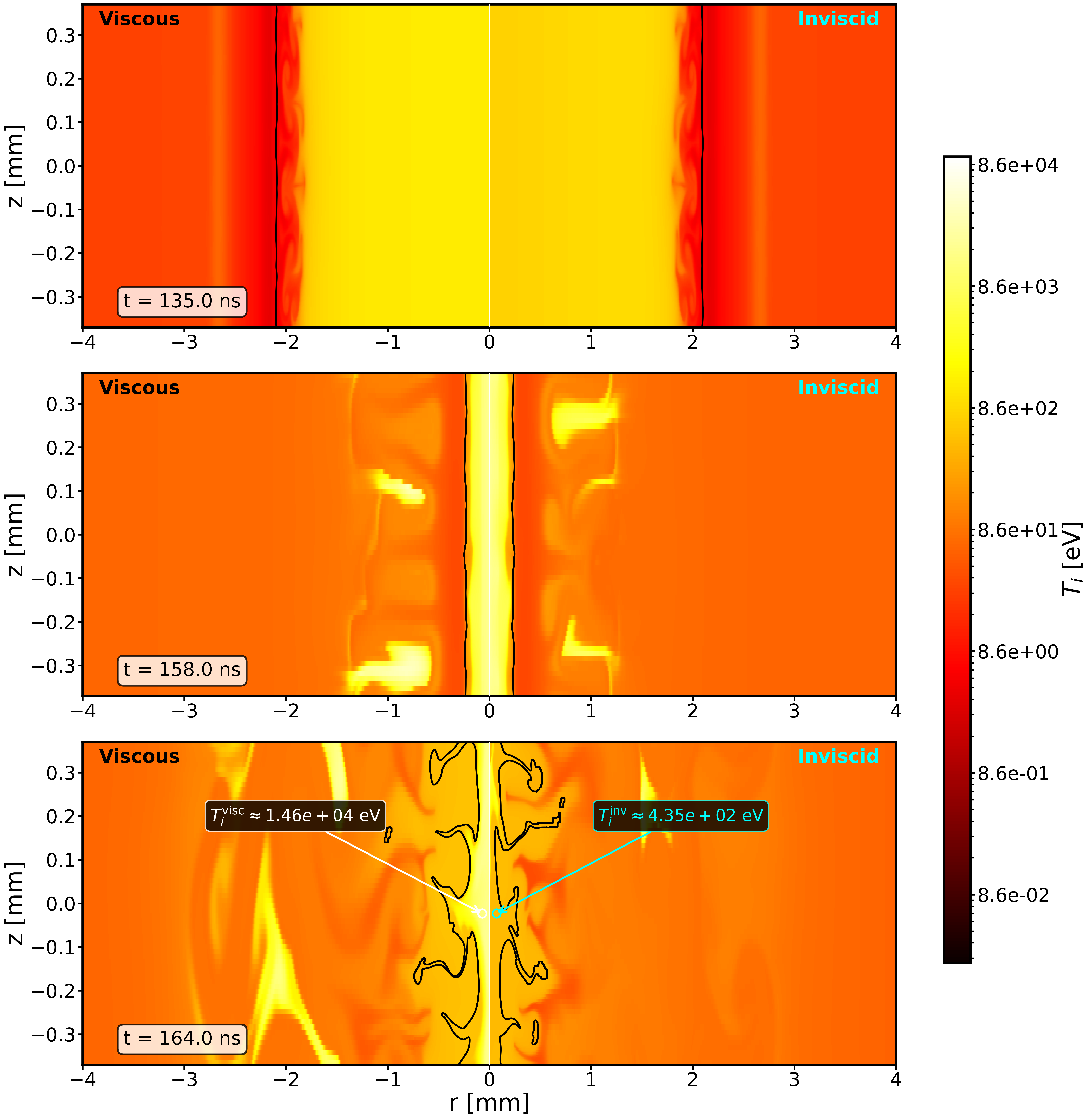}
    \caption{
        Ion temperature evolution for the viscous (left half of each panel)
        and inviscid (right half) simulations, shown at
        \(t = 135\), 158, and 164 ns (top to bottom). The fuel temperature
        remains appreciably higher in the viscous case throughout the
        implosion, with the largest contrast at stagnation where a hot,
        axially extended fuel column is preserved in the viscous run while
        the inviscid hot spot has cooled and mixed with colder liner
        material. In the bottom panel, a digital probe placed near
        \(r \approx 0\), \(z \approx 0\) (white and cyan markers with arrows)
        reports a central ion temperature that is more than an order of
        magnitude higher in the viscous case than in the inviscid case,
        illustrating the strong viscous heating of the hot spot. The black contours represent the aluminum boundary.
    }
    \label{fig:maglif_tion_evolution}
\end{figure}

Figure~\ref{fig:maglif_tion_evolution} shows the evolution of the ion temperature \(T_i\) for the same three late-time snapshots as in the density evolution, at \(t = 135\), 158, and 164 ns. Using the same mirrored visualization, the viscous solution occupies the left half of each panel while the inviscid solution occupies the right half. At all times the fuel temperature is systematically higher in the viscous case, both along the axis and near the fuel–liner interface. To quantify this contrast, we place a digital ``thermometer'' probe at \(r \approx 0\) and \(z \approx 0\) in the stagnation frame (bottom panel). The probe reads \(T_i^{\mathrm{visc}} \approx 1.46\times10^{4}\,\mathrm{eV}\) in the viscous run and \(T_i^{\mathrm{inv}} \approx 4.35\times10^{2}\,\mathrm{eV}\) in the inviscid run, so the central fuel temperature in the viscous case is more than an order of magnitude higher compared with the inviscid case. By stagnation (\(t = 164\) ns) the viscous run maintains a hot, axially extended column of fuel on axis, whereas the inviscid run has cooled substantially as cold liner material penetrates and mixes into the hot spot. This behavior is consistent with the increased temperature in the viscous simulations discussed in Section~\ref{sec:pool_heated}.

\begin{figure}[htbp]
    \centering
    \includegraphics[width=\columnwidth]{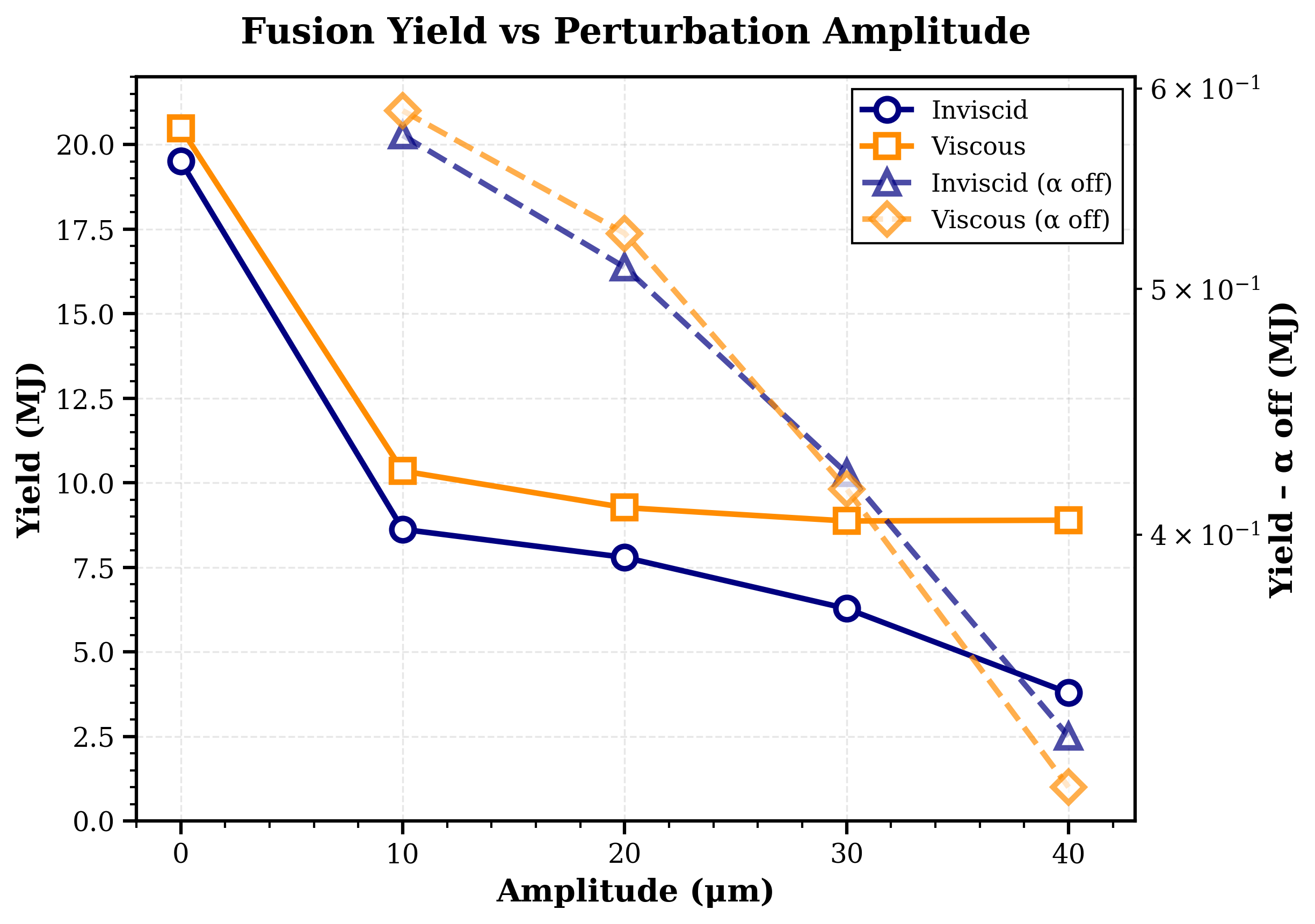}
    \caption{Fusion yield as a function of initial perturbation amplitude for viscous and inviscid simulations at a fixed wavelength of 200~$\mu$m. Solid lines with markers show simulations with alpha particle deposition enabled; dashed lines show simulations with alpha deposition disabled (shown on log scale). The viscous simulations consistently produce higher yields across all perturbation amplitudes. Notably, the yield in the viscous case asymptotes at large amplitudes, whereas the inviscid yield decreases monotonically. The maximum yield preservation of approximately 134\% occurs at an amplitude of 40~$\mu$m. Without alpha heating, yields are dramatically lower ($\sim$0.3--0.6 MJ) and the viscous-inviscid differences are negligible.}
    \label{fig:yield}
\end{figure}

Figure~\ref{fig:yield} presents the fusion yield as a function of initial perturbation amplitude for both viscous and inviscid simulations. Several key observations emerge from this comparison. First, the viscous simulations consistently produce higher yields across all perturbation amplitudes studied (with $\alpha$ on). Second, the yield curves exhibit qualitatively different behavior: while the inviscid yield decreases monotonically with increasing perturbation amplitude, the viscous yield appears to asymptote at larger amplitudes, suggesting that viscosity provides increasingly effective stabilization as instabilities grow more severe. Third, even in the absence of explicitly seeded perturbations (amplitude = 0), the viscous simulation produces a modestly higher yield. We attribute this primarily to viscous heating of the compressional flows during the implosion, as discussed earlier: the viscous heating term provides an additional heating mechanism that operates even in purely radial flows without any instability development.

The yield preservation is substantial, reaching approximately 134\% at the largest perturbation amplitude studied (40~$\mu$m). To isolate the role of viscosity from alpha heating feedback, we also performed simulations with alpha particle deposition disabled (dashed lines in Figure~\ref{fig:yield}). Without alpha heating, the yields are dramatically lower ($\sim$0.3--0.6 MJ compared to $\sim$4--21 MJ), and the differences between the viscous and inviscid cases are negligible. The much larger yield differences in the alpha-on cases suggest that viscosity-induced improvements in hot spot integrity are amplified by the positive feedback of alpha heating: better confinement leads to more alpha deposition, which further increases temperature (which further increases the effect of viscosity) and fusion rate. Analysis of neutron-averaged quantities (Table~\ref{tab:navg_quantities}) reveals that both the pressure and density in the burn region are consistently higher in simulations with viscosity included, with particularly striking enhancements at the largest perturbation amplitude: $+35.7\%$ in pressure and $+25.1\%$ in density. These substantial increases indicate that viscous damping of deceleration-phase instabilities leads to improved fuel compression and confinement. Interestingly, the neutron-averaged ion temperature is slightly \emph{lower} in the viscous cases (by $\sim$0.1--1.9\%), which appears to contradict our earlier finding that viscosity produces higher temperatures on average. However, this comparison is restricted to the neutron-producing region, whereas the temperature enhancements observed in Section~\ref{sec:pool_heated} were distributed throughout the fuel volume; the neutron-averaged temperature is weighted toward the hottest regions where viscous heating may be less significant relative to compressional and alpha heating. Given the small magnitude of this difference compared to the pressure and density enhancements, we leave detailed investigation of this effect to future work.

\begin{table*}[htbp]
\centering
\caption{Neutron-averaged quantities for viscous and inviscid simulations at varying perturbation amplitudes. Percentage differences compare viscous to inviscid cases, with green indicating increases and red indicating decreases. Color intensity reflects the magnitude of the difference.}
\label{tab:navg_quantities}
\begin{tabular}{cccccccc}
\hline
\hline
Viscosity & Amplitude & $\langle T_i \rangle_n$ & $\langle P \rangle_n$ & $\langle \rho \rangle_n$ & $\Delta T_i$ & $\Delta P$ & $\Delta \rho$ \\
 & ($\mu$m) & (keV) & (Gbar) & (g/cm$^3$) & (\%) & (\%) & (\%) \\
\hline
N & 10 & 13.86 & 36.60 & 4.57 & --- & --- & --- \\
N & 20 & 14.08 & 35.91 & 4.46 & --- & --- & --- \\
N & 30 & 13.50 & 36.56 & 4.68 & --- & --- & --- \\
N & 40 & 12.26 & 33.94 & 5.05 & --- & --- & --- \\
\hline
Y & 10 & 13.60 & 36.52 & 4.56 & \cellcolor{red!15}{$-1.9$} & \cellcolor{red!5}{$-0.2$} & \cellcolor{red!5}{$-0.2$} \\
Y & 20 & 13.93 & 39.27 & 4.73 & \cellcolor{red!10}{$-1.1$} & \cellcolor{green!20}{$+9.4$} & \cellcolor{green!15}{$+6.1$} \\
Y & 30 & 13.36 & 39.27 & 4.86 & \cellcolor{red!10}{$-1.0$} & \cellcolor{green!18}{$+7.4$} & \cellcolor{green!10}{$+3.8$} \\
Y & 40 & 12.25 & 46.07 & 6.32 & \cellcolor{red!5}{$-0.1$} & \cellcolor{green!60}{$+35.7$} & \cellcolor{green!50}{$+25.1$} \\
\hline
\hline
\end{tabular}
\end{table*}

While a comprehensive analysis of the spectral dependence of viscous stabilization---including sweeps over wavenumber and amplitude-to-wavelength ratio to quantify the effect on instability growth rates---lies beyond the scope of this work, the results presented here clearly demonstrate that magnetized viscosity has a non-negligible and beneficial impact on MagLIF implosion performance. These findings motivate the inclusion of anisotropic viscous transport in predictive simulations of MagLIF and related magnetized ICF concepts.

\section{Conclusion}
\label{sec:conclusion}

We have presented the first implementation of the full Braginskii magnetized viscosity tensor in an implicit solver within a multiphysics radiation-magnetohydrodynamics framework. The implementation handles arbitrary magnetic field orientations by constructing the complete anisotropic stress tensor from five independent viscosity coefficients, each governing distinct momentum transport processes ranging from field-parallel compression to gyroviscous effects arising from finite ion Larmor radius physics. Our implicit backward Euler treatment of velocity diffusion eliminates restrictive viscous CFL constraints that would otherwise dominate timesteps in high-viscosity regions, enabling efficient simulation of the strongly magnetized regime characteristic of MagLIF plasmas where the parallel viscosity coefficient can exceed perpendicular coefficients by many orders of magnitude.

The implementation has been verified through four complementary test cases of increasing complexity. Comparison with an approximate analytical solution for velocity diffusion in the case where the magnetic field and velocity gradients are aligned shows good agreement. Direct verification against Braginskii’s original analytic form for a purely out-of-plane (azimuthal) magnetic field confirms that the general arbitrary-field implementation correctly reduces to the expected limiting case. In addition, Method of Manufactured Solutions testing for a fully three-dimensional magnetic-field topology with all three field components active demonstrates that the solver converges at the expected formal order, namely second order in space and first order in time. Finally, comparison against semi-analytic magnetized viscous shock profiles shows excellent agreement, confirming the correct coupling between the Braginskii viscosity module and the hydrodynamics solver. Taken together, these results provide strong evidence that the implementation is correct and that the discretization errors decrease as expected under spatial and temporal refinement.

Application to MagLIF-relevant configurations reveals that magnetized viscosity has a pronounced and beneficial effect on implosion dynamics. In simulations of pool-heated targets, viscosity effectively damps vortical structures that would otherwise persist in inviscid calculations, converting the kinetic energy stored in these rotational motions into thermal energy of the plasma. The resulting temperature enhancement reaches approximately 10\% of peak values through the combined mechanisms of direct viscous heating and the dissipation of vortical kinetic energy. In simplified MagLIF configurations with seeded Rayleigh-Taylor perturbations, viscous simulations consistently produce higher fusion yields across all perturbation amplitudes studied, with maximum yield enhancements reaching 134\% at the largest perturbations. The qualitatively different behavior of the yield curves---asymptoting at large amplitudes in the viscous case versus monotonically decreasing in the inviscid case---suggests that viscosity provides increasingly effective stabilization as instabilities grow more severe.

These results establish magnetized viscosity as a non-negligible physical mechanism that should be included in predictive simulations of MagLIF and related magnetized inertial confinement fusion concepts. The demonstrated yield improvements and enhanced fuel compression highlight the importance of capturing anisotropic transport physics for accurate performance predictions. Future work could include extending this capability to three-dimensional simulations, and conducting systematic parameter studies to quantify viscous stabilization across the full space of perturbation wavelengths and amplitudes relevant to MagLIF target design. As Pacific Fusion advances toward its 60 MA Demonstration System, the simulation capabilities developed here will contribute to the validated modeling framework essential for achieving facility gain and advancing toward commercial fusion energy.


\begin{acknowledgments}
This work was funded by Pacific Fusion's 2025 Summer Internship Program.
\end{acknowledgments}

\section*{References}
\bibliography{References}
\end{document}